# A Theoretical Framework for Physics Education Research: Modeling Student Thinking

Edward F. Redish

*University of Maryland – College Park, MD USA*

*Summary.* – Education is a goal-oriented field. But if we want to treat education scientifically so we can accumulate, evaluate, and refine what we learn, then we must develop a theoretical framework that is strongly rooted in objective observations and through which different theoretical models of student thinking can be compared. Much that is known in the behavioral sciences is robust and observationally based. In this paper, I draw from a variety of fields ranging from neuroscience to sociolinguistics to propose an over-arching theoretical framework that allows us to both make sense of what we see in the classroom and to compare a variety of specific theoretical approaches. My synthesis is organized around an analysis of the individual's cognition and how it interacts with the environment. This leads to a two level system, a knowledge-structure level where associational patterns dominate, and a control-structure level where one can describe expectations and epistemology. For each level, I sketch some plausible starting models for student thinking and learning in physics and give examples of how a theoretical orientation can affect instruction and research.

## 1 – Motivation and Introduction

### 1.1 : Identifying a Theoretical Framework

Education research is an applied field. As educators, we want to understand how teaching and learning works in order to be able to teach our students more effectively. As scientists, we would like to do this using a scientific approach that combines observation, analysis, and synthesis like the one that has been so effective in helping us make sense of the physical world. Such a synthesis helps transform a collection of independent "facts" into a coherent science, capable of evaluating, refining, and making sense of our accumulated experimental data. However, education research differs from traditional physics research in that in education our goals often dominate our view of the system we are trying to understand.[*] We never want to lose sight of our goals, but we very much want to treat our system scientifically in order to understand how it functions. Indeed, if we better understand how the system functions, we can better articulate and refine our goals. Moreover, we might be able to explain just what it is that highly successful teachers do and transform what is presently an art into a teachable science. At present, despite a few researchers who discuss theoretical frames, research and development in education is strongly dominated by observation and by direct educational goals: "What do we have to do to get our students to learn more effectively?"

But science is not just a collection of observations: "These students do thus in these circumstances."  In building a science we depend heavily on the idea of *mechanism* – describing the behavior of a system in terms of a small number of objects or variables. Such a description tells us what we are talking about and how we are going to think about it. I refer to this choice of objects and variables as our *ontology*.

Seeking mechanism is not just reductionism, though reductionism (description of a system in terms of the behavior of fundamental constituent parts and their interactions) can often provide mechanism. Sometimes in physics we create *collective variables* – like Cooper pairs, phonons, pressure, or temperature.[†] We try to isolate "what matters" in describing a physical system so as to produce an optimal description – one in which the system and its behavior are described by a minimal number of concepts and one in which the complex behavior of the system arises from combinations and elaborations of the simple structures and their interactions.

If we are going to try to study education using the tools and methods of science, we need to develop a *theoretical framework* – a shared language and shared assumptions that can both guide and allow us to compare different approaches and ways of thinking. An example of a theoretical framework in atomic, molecular, and condensed matter physics is the theory describing matter as electrons and nuclei satisfying a many-body non-relativistic Schrödinger equation with Coulomb and (first order) radiative electromagnetic interactions. Although this framework is widely believed to provide highly accurate descriptions of atoms, molecules, and matter, it can in actual fact only be used to calculate the properties of a very small number of systems (hydrogen, helium, the $H_2^+$ ion, etc.).

Calculating more complex systems requires a *model* – a starting point for the description of the complex system that assumes a simplified structure for the behavior of most of the particles in the system. The atomic shell model is one example. The Bloch waves and Fermi surface model of electrons in a crystal is another. Each of these models is constrained by and guided by the over-arching theoretical framework and their imbedding in that framework may permit the calculation of corrections to the model or the calculation of the model's phenomenological parameters (if it has any).

---

[*]   We should not ignore the fact that a practical goal is implicitly imbedded in much of traditional science: the goal of learning how to control our environment. There is a continual tension between basic and applied science that arises from the inevitable imbedding of science in a social context.

[†]   What is a collective variable and what is fundamental can change depending on our theoretical frame. An electric field is fundamental in a classical picture. In a photon picture it is a collective variable.

In physics, we tend to refer to theoretical structures that address a fairly narrow range of phenomena, such as the low-lying energy levels of atoms or nuclei, as *models*. We call the broader dynamical framework in which these models are imbedded as *theories*. In education (and in cognitive science), the tendency is to refer to the former as theories (e.g., the theory of small-group social interaction, or the modular theory of students' senses of physical phenomena) and the latter as a *theoretical framework*. For this paper, I will use the physics terminology, but modify "theory" to "theoretical framework" when I want to stress the incompleteness of the structure.

In this paper, I propose the outline of a few components of a theory appropriate for thinking about how teen-agers and young adults learn physics. Into this framework I collect and propose some appropriate models in the hopes of encouraging a dialog on theoretical issues. My goal is to try to help the community begin to establish a few foothold ideas by seeking common ground among distinct models. Our theoretical frame and the models I describe in what follows synthesize and extend a number of good ideas that have been known both to researchers and some teachers for many years. A major part of trying to develop a theoretical structure is to be able to go beyond the "tips and guidelines" that successful teachers and researchers provide us and to see how to fit these suggestions into a broader and more coherent structure that can be explained and transferred to others.

*1.2 Constructing a Theoretical Framework*

Where in the complex system of students in a classroom should we begin to construct a theoretical framework? The education of a student is an immensely complex issue. Each student is an individual with a complex mental structure and responses. Those mental structures have been formed by the interaction of the individual's genetic possibility with their environmental development. In addition, each individual lives in many cultures and is educated in many social environments that play a major role in what the student learns (and does not learn).

Three broad issues play major roles in learning, even for a single individual: the development of the individual's mental system, the behavior and functioning of the individual, and the interaction of the individual to respond to and help create a social environment. Each of these issues has been studied extensively and much is known. In this overview, I choose to focus on what appears to me to be the central issue: the behavior and functioning of individual adults – high school and college students – particularly in the context of the learning of science (and of that, particularly learning physics, from which most of my examples will be drawn). Developmental issues, while playing a role in establishing the structures observed in the individual, are indirectly related to the educational issues we are interested in here. Socio-cultural issues, however, play a critical role. Every adult's thinking processes have been shaped by being raised within a culture and these processes both respond to and shape the cultural environments in which individuals find themselves. It is possible – and valuable – to view the individual as part of a social system of a variety of grain sizes. This adds an further complexity to the issue of understanding how an individual thinks. For a discussion of some of these issues, see Otero's paper in this volume and the references there. [1]

In this paper, I only address socio-cultural issues "from the inside" – that is, from the point of view of the individual and how the individual's cognition responds to both the socio-cultural and physical environments. Even if we are primarily interested in socio-cultural phenomena, what is learned from the individual cognitive perspective should be useful. When considering a system of objects, it is often helpful to understand the character and behavior of the individual objects in the system.

If we restrict our theoretical framework to the cognition of the adult individual and how he responds to his physical and social environment, where do we begin? We are trying to describe one of the most complex systems known on earth: human behavior. This is not rocket science – it's MUCH harder. To see how hard, we can describe the system in physics terms: It is a strongly interacting many-body system in which observations change the system in uncontrollable ways. We therefore want to be modest in what we try to achieve at this stage, but to rely as much as possible on what has been learned. Since phenomenological modeling of human behavior in education and psychology is sometimes "all over the lot" I rely heavily on a triangulation through results in multiple fields: fundamental cognitive research, neuroscience, and research on real people doing real tasks in real situations.[*] This last involves many disciplines including, educational research, ethology, sociology, anthropology, and sociolinguistics. I organize this into three levels: neuroscience, cognitive science, and the phenomenological observational sciences of human behavior.

As a physicist, I naturally tend to be a reductionist: I want to be able to conceive of mechanisms underlying the phenomena I describe even if the connection is difficult or somewhat obscure. Since the brain is composed of biological components – particularly neurons – the study of the mechanical functioning of this system strikes me as having relevance, even if we are far from understanding how thought and understanding arise from biological processes. Neuroscientists have begun to build an understanding of the biological mechanisms that underlie some aspects of human behavior – analogous to building a statistical mechanics of the collective variables determined by the psychological phenomenologists. [2][3][4] I review some basic results of neuroscience in section 2.

But it is not appropriate at this stage (or perhaps at any stage in the foreseeable future) to attempt a reductionist description of human behavior.[†] What we want is to construct a mid-level set of collective variables – a *mesoscopic* thermodynamics of thinking – that provides a useful ontology for constructing mechanisms.

Fundamental cognitive research attempts to investigate the underlying ontology and mechanisms of the human mind – to "carve the mind at its joints."[5][6][7] Since the mind is extremely complex and is often able to compensate for deficiencies in one area by repurposing or reinterpreting data from other areas, there are rarely single identifiable causes for any given response. To attempt to isolate mechanisms, research psychologists create experiments that may appear highly contrived, such as ones measuring time delays of milliseconds in responses, saccadic eye motions, or a subject's ability to recall nonsense syllables. As a physicist, I recognize these kinds of experiments and feel quite comfortable with their design, in principle, if not in detail; they are *zero-friction experiments*. In physics, in the effort to isolate mechanisms we often go to great efforts to suppress phenomena that are present in every real-world situation and that may play a critical role in what actually happens. Our enhanced understanding of underlying mechanisms allows us to re-interpret what we see in the real world in a more coherent fashion. But we have to be very careful to "put the friction back" before drawing practical conclusions. A small number of elements from the large body of cognitive science results are summarized in section 3.

---

[*]   This kind of research is referred to as *ecological* in psychology.

[†]   To keep reductionism in perspective, there are $\sim 10^5$ neurons and $\sim 10^9$ synapses per cubic millimeter of brain tissue. Furthermore, the system cannot be treated statistically since there is considerable organization – though not necessarily on the neuron by neuron level. [4]

In order to understand how students build new knowledge and how students respond to different classroom contexts, I use information from these two fundamental sciences to categorize behavior into two broad areas: association and control. In each of these areas I outline a theoretical framework and then discuss a few of the models that have been proposed that fit nicely in this structure and that are relevant for the teaching and learning of physics. In section 4 I discuss associational patterns: knowledge structures, cognitive resources, and their patterns of association. In section 5 I discuss control: epistemology, expectations, and framing. In section 6 I consider applications of this theoretical structure to instruction and to research. Section 7 discusses some conclusions. Since I am building by synthesis by combining many different areas of research, terminology can be a problem. Different areas of research use the same term in different ways (as, indeed, do competing researchers in the same research area.). To provide some concreteness and clarification, in section 8 I provide of glossary of terms.

## 2 – The starting point: a foothold in neuroscience

Our starting point in building our theoretical framework is the assumption that underlies the operation of "normal science."[8]

> *Principle 1: (Working hypothesis)* All phenomena are describable as arising from the fundamental physical objects and laws that we know.

Of course, we don't know all physical laws and objects, but we know a lot. Our principle 1 suggests that we assume that there is no "new physics" until we are forced to do so by the data. Thus, we should not assume *ab initio* that organic chemistry requires a fundamentally different treatment of atoms in molecules than inorganic chemistry. In the case of cognition, the principle of "trying to do normal science first" says that we should assume:

> *Principle 2:* All cognition takes place as a result of the functioning of neurons in the individual's brain.

This means that we will not assume a "mind" or "spirit" that is somehow superposed on and different from the functioning of a brain's neurons. [9] In the spirit of normal science we will hold to this assumption until forced to modify it by extensive data.[*]

### 2.1 The basic ideas of neuroscience

At present, neural research suggests that knowledge and learning are carried by the set of neurons of an individual's brain and their connections. Neurons are cells that have long cylindrical protuberances (dendrites and axons) that are electrically active and that connect to other neurons (and to sensors and muscles) at their ends (synapses). (See figure 1.)

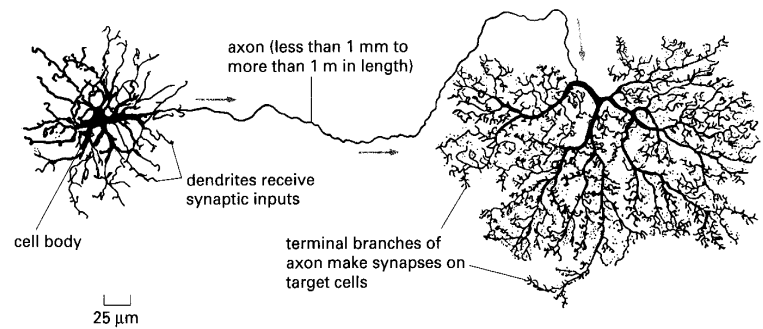

*Fig. 1: Neurons and neural connections. From [11], p. 1228 (courtesy, B. Alberts).*

In $0^{th}$ approximation, dendrites and axons are basically long thin cylindrical capacitors. They maintain a potential difference across their inner and outer membranes. When the cell is activated, the cylinder discharges axially in a small region and this region of localized discharge runs down the cylinder. This phenomenon is called an *action potential* and requires active electrical mechanisms to produce it. As shown in figure 2, an activated cell produces a chain of action potentials. The information carried by this signal appears to be contained mostly in the rate at which these pulses are produced. ([2] but see also [12])

Actual thought and cognition occurs when neurons are activated. We don't really need to know much about neurons and the complexity of their functioning, but there are a few basic "foothold" ideas that constrain the kinds of models we can build and that give us a sense of mechanism about cognitive processes. [2]

*Principle 3: Neuronal foothold principles:*

3.1.  Neurons connect to each other.
3.2.  Neurons send information to each other via pulse trains when they are activated.

---

[*] There are many examples of cognitive phenomena that could possibly be seen as "emergent" phenomena – behaviors that are not visible when viewed from the system's component parts. Consciousness is the most obvious. However, see Dennett [10] and Damasio [9].

3.3. Neurons may be in various stages of activation.

3.4. Multiple neurons can link to a single neuron.

3.5. Signals from one or more neurons can result in the activation of linked neurons.

3.6. Neural connections can enhance or inhibit other neural connections.

3.7. Information flows both from a set of neurons (e.g., sensory neurons) to processing neurons (*feed-forward*) and back (*feedback*).

3.8. Learning appears to be associated with the growth of connections (synapses) between neurons.

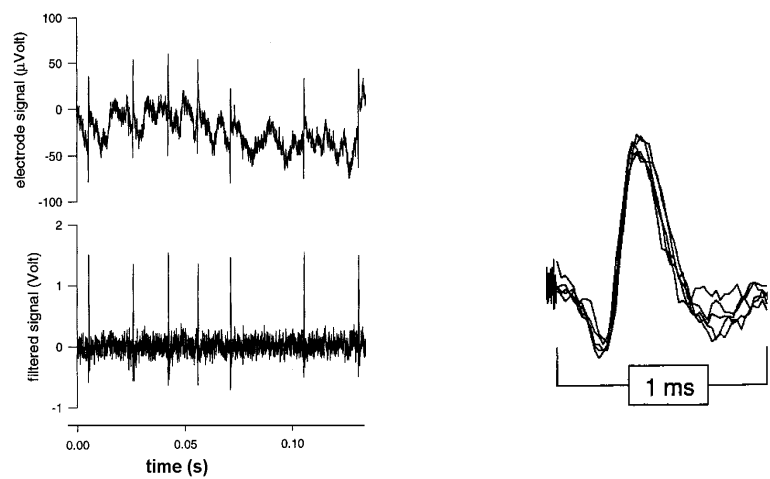

*Fig. 2. Top left: potential difference between the surrounding fluid and a point near the cell in the brain of a fly. Bottom left: same signal filtered to remove low frequencies. Right: Voltage pattern of 5 action potentials overlaid. (From [12] p. 5, courtesy W. Bialek.)*

The fact that activation of one (or more) neurons can lead to activation of other neurons has profound implications. This leads to the idea of *association*, one of our fundamental tools for making sense of the cognitive response. The ideas that activation can be either inhibiting or enhancing and that the neural system contains considerable feedback underlies the concept of *control*, my second fundamental categorization of cognitive processing.

Neuroscience has much interesting to say about neural development. One important point for understanding infants and very young children is that the brain is first built with far more neurons than appear to be needed. As a result of experience, new connections are made, but many cells die off. [13] [14] One implication is that early experience is extremely important. A kitten whose vision in one eye is blocked for the critical few early weeks never learns to see out of that eye, even though the eye may be fully functional.* A developmental point that may be relevant for young adults is myelination. Long axons develop a sheath that speeds up the transmission of signals along the axon dramatically. Some of the axons in the brain do not myelinate until well after puberty, suggesting that it is reasonable to infer that some cognitive functions could be late in developing.[*] [15] Interesting as this is, an extensive discussion of neuroscience and development is beyond the scope of this paper. Instead, we turn to consider how understanding some of what has been learned in neuroscience helps us understand cognition.

---

* This result is species specific, and does not hold, for example, for ferrets.

### 2.2 Fine-grained constructivism and resources

The fundamental results of neuroscience have inspired some approaches to applied cognition. Some researchers, both cognitive scientists [16] [17] and educational researchers [18], build very specific and detailed models of specific cognitive tasks. Another group creates computer programs to try to explicate specific steps and tasks necessary to complete a particular cognitive activity. [19] I will refer to these approaches collectively as *connectionism*.[†] Although connectionist approaches are interesting and may someday tell us something useful about cognition, at their present stage of development they seem rather far from what we need to understand what is happening in our classrooms.

These approaches are trying to build microscopic theory of cognition, but what we need for studying education is to identify intermediate scale categories, collective variables, and principles that can guide our understanding of what we see and do in real educational environments – a *mesoscopic* theory. These structures should be at a fine enough grain size that they allow us to understand the structure of everyday thinking, but not so fine as to require a highly complex analysis for every statement anyone makes. We need to look for fundamental tools with general implications that arise from our assumption of principle 2 and the assumed characteristics of neural behavior. In some sense, the neuronal model provides us metaphors for the structure of thought that help us develop a way of thinking about thought.[‡]

Even our simple neural footholds listed under principle 3 lead us to a number of useful fundamental ideas. The first is what I like to call "neural solipsism imbedded in an external reality."

> *Principle 4:* There is a real world out there and every individual creates his or her own internal interpretation of that world based on sensory input.

Another way of saying this is: We each live in our own virtual reality, but we try to make that virtual reality as good an approximation to the true reality as possible. Culture in general and science in particular is a process we create to try to help each of us, alone in our own mental

---

[*]   Note, however, that Fuster comments ([2] p. 64) that "nowhere in the nervous system is myelination a precondition for axonal involvement as some level of function." It might, however, affect the extent to which compilation (see section 3.1.1) can take place.

[†]   For a clear summary and description of the strengths and weaknesses of connectionism, see [20], chapter 6.

[‡]   Another useful metaphor for thought that meshes well with the neural is that of coupled non-linear oscillators. See Euler's talk in this volume, in particular, the discussion of entrainment. [21]

space, to use our interactions with others in a community to improve our maps of what's out there.

Our over-arching interest as education researchers is in understanding what reality individuals have created and how they modify that view of reality as they acquire new knowledge. A fundamental hypothesis is:

> *Principle 5:* New knowledge is built on a base of existing knowledge by building new links and suppressing old ones.

Note that this principle only makes sense if we restrict our consideration to adults and older children. It is obvious to any parent who has watched an infant learn to cope with the world that at least some knowledge, if not "hard-wired" at birth, is "set up" to wire automatically given appropriate early environmental experiences.[*]

Many educational researchers will recognize this as the neurological translation of the fundamental principle of *constructivism* – that individuals build their new knowledge on a base of their existing knowledge. What we are particularly interested in is *fine-grained constructivism*. We want to analyzing knowledge into more fundamental components in order to understand <u>how</u> that construction takes place. Even such basic neural processes as perception are not just simple connections. Inputs are highly transformed and processed. [21] Analogously, new knowledge can be created from old by extension, elaboration, and transformation. Creation of new knowledge depends on activation of resources that are already there – not just for recall but for processing, transforming, and recombining.

I will refer to this overall structure as the *resources* approach. [23] In this approach, we want to answer the question: When a student responds to an instructional environment to build new knowledge, what existing resources are activated and how are they used? Our goal is to develop a sufficient understanding of the cognitive structure of knowledge building to be able to predict what environments are likely to be more effective for more students. Our theoretical framework says resources should exist. A more detailed cognitive theory is required to specify what those resources are. Let's next consider what has been learned from cognitive science that can help us.

## 3  – Cognitive mechanisms: Association and control

The information we have from the neural level is too detailed and too limited to provide, by itself, much direct help in our classroom environments. We need to supplement what we learn at the neural level by observations of individual's behavior in controlled and natural environments.[†]

---

[*] Piaget calls this latter situation *epigenetic*. Fuster refers to pre-programmed structural knowledge as *phyletic memory*. [2] See the collected works of Piaget or [22].

[†] In addition to the study of normal individuals, a significant point of contact between neuroscience and cognitive science is the study of *lesions*. In this area of research, the cognitive behavior of individuals with damage to specific parts of their brains is studied. This gives valuable ontological information: how cognitive behavior is parsed in the brain. [24] [25]

When we observe the actual behavior of individuals, we expect that observed behavior involves the activation of large numbers of neurons. Nonetheless, it seems plausible that even groups of neurons will show the structures that are demonstrated at the level of individual neurons. Three fundamental concepts we inherit from neuroscience are

- activation,
- association,
- enhancement/inhibition.

At the cognitive level, analogous structures recognized by cognitive researchers include

- recall and priming,
- linking, compilation, and spreading activation,
- control and executive function.

I discuss each briefly from the point of view of cognitive studies before turning to a detailed discussion of some theories for helping us understand learning and instruction.

### 3.1  The structure of memory: putting stuff in (learning) and getting it out (recall)

To understand learning, we must understand memory – how information is stored in the brain. Modern cognitive science now has complex and detailed structural information about how memory works. [2] [6] [26]

Over the years, there have been many attempts to find structures in human memory. Two groups of observations have been critical in building what we know: the observation of amnesiacs and controlled experiments with normal individuals. Amnesia often results from brain injury. In many cases, the individual loses some memory capabilities while retaining others.[*]  In controlled experiments, psychologists have been able to map in great detail the timing and character of certain (simple) parts of the memory system. The basic structure of memory that many psychologists and neuroscientists currently accept is illustrated in figure 3. We don't really need all of the structures in this complex diagram, but I include it to give you an idea of what mental structures are reasonably well established by cognitive- and neuro-scientists. If you are interested in more information about the elements of this diagram, consult [2] and [26].

Input from the external world comes in through the senses and is processed in a fast sensory pre-processing system. The output of this system is then combined in a short-term or working memory in which the processed sensory information is mixed with information from a long-term store. An executive controls what is done with the mixed data through the mechanisms of paying attention and conscious thought.

---

[*]  In a famous case (H.M., see [27]), a patient, part of whose brain had been surgically removed, lost the ability to create long-term episodic memories. He could conduct conversations but not remember them even a few minutes later but he was able to remember events before his surgery and learn new motor skills (but not remember learning them).

Transfer of results created in working memory into long-term memory is not immediate but requires repetition and significant time, often days or weeks.[*] In our neural model, this is plausible since we expect that formation of memories may actually require growth of new synapses.

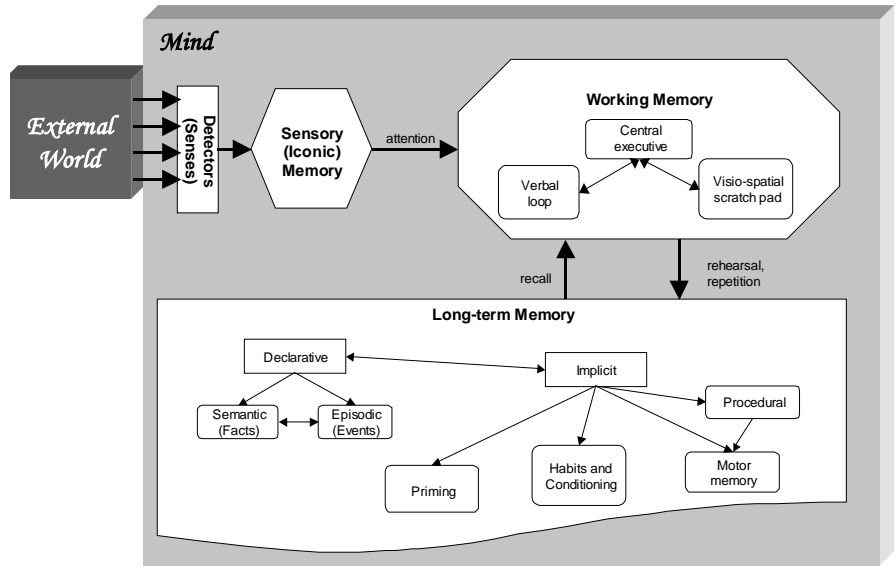

*Fig. 3: A map of the structure of human memory. Adapted from [2] and [26].*

Long-term memory can be divided into two parts, *declarative memory* and *implicit memory*. Declarative memory contains knowledge that can be articulated, such as facts and consciously followed procedures (semantic memory), and memories of events and situations (episodic memory). Implicit memory concerns procedural and motor memories that are not articulated – like knowing how to run or ride a bicycle – as well as habits and conditioned responses.

It is clear from all the different things that people can do that require memory that memory is a highly complex and structured phenomenon. Fortunately, we only need to understand a small part of the structure to get started in learning more about how to teach effectively. Six foothold ideas from cognitive science are:

*Principle 6:* Cognitive foothold principles*:*

   6.1.   Memory has two functionally distinct components: working (or short term) memory and long-term memory.

---

[*] Highly emotional events are processed through different systems and can occur much more quickly than day-to-day learning.

6.2. Working memory can only handle a small number of data blocks. It is labile, often lasting only a few seconds without specific activities to prolong it.

6.3. Long-term memory contains a vast quantity of facts, data, and rules for how to use and process them (declarative and implicit memory). It is highly stable and can store data for decades.

6.4. Getting information from working to long-term memory requires repetition.

6.5. Getting information from long-term memory to working memory may be difficult and time consuming.

The details of this structure are still somewhat controversial. For example, it is clear that working memory has distinct verbal and visual parts. Subjects can carry out simple verbal and visual tasks simultaneously without interference, but two verbal or two visual tasks interfere with each other.[*] [5] [6] The independence of these two memory modules can be illustrated by simple experiments in which subjects try to remember lists of words while doing other tasks. These structures will have implications for us when we discuss the use of various representations in displaying physical information.[†] Recent evidence suggests there may be other independent components to working memory. [29]

The point 6.5 in our cognitive foothold list turns out to have great importance for us. If we "know something," that is, if it is present in our long term memory, we might not be able to access it immediately. We have all had the experience of meeting someone we know well out of the normal context in which we know them and being unable to recall their name. What is needed to activate the memory is a *chain of associations*. This is discussed in section 3.2.

### 3.1.1   *Chunking*

Working memory is basically the part of the brain activated when we are thinking about something. Working memory is fast but can only handle a fairly small number of "units" or "chunks" at one time. Early experiments by Miller [30] suggested that the number was "7±2". More recent experiments indicate that this is an oversimplification. Semantic memory, for example, typically seems to be able to hold only "3±1" items.

We cannot understand these numbers until we ask "What do they mean by a chunk?" Miller's experiments involved strings of numbers, letters, or words. But clearly people can construct very large arguments!  If I had to write out everything that is contained in the proof of a theorem in quantum field theory it would take hundreds, perhaps thousands, of pages. The key, of course, is that I don't write out or explicate everything when I am thinking about a piece of the proof. My knowledge is combined into hierarchies of blocks (or chunks) that I can work with even with my limited short-term processing ability.

---

[*]   In the cognitive and neuroscience literatures these are referred to as the *auditory* or *phonological loop* and the *visuo-spatial sketchpad*.

[†]   As a simple example of interference, try taking your pulse while whistling a song. This is difficult, but we seem to have no trouble keeping track of a song's words and music at the same time.

To see how this works, try multiplying a pair of two-digit numbers in your head. Stop and try it before going on. Did you keep all of the parts in mind at once? Or did you create subtotals, forgetting about the numbers that went into building them? Most of us who work frequently with numbers can multiply two-digit numbers without "putting stuff away in storage." Now try doing it with a pair of three-digit numbers. The difference between having fewer than "7±2" objects to work with and significantly more becomes painfully clear.

Another example in which you can see the structure of working memory in your own head is to try to memorize the following string of numbers:

3     5     2     9     7     4     3     1     0     4     8     5

Look at it, read it aloud to yourself or have someone read it aloud to you, look away for ten seconds and try to write the string down without looking at it. How did you do? Most people given this task will get some right at the beginning, some right at the end, and do very badly in the middle. Now try the same task with the following string

1     7     7     6     1     8     6     5     1     9     4     1

If you are an American <u>and</u> if you noticed the pattern (try grouping the numbers in blocks of four) you are likely to have no trouble getting them all correct – even a week later.

The groups of four numbers in the second string are "chunks" – each string of four numbers is associated with a year, not seen as four independent numbers. The interesting thing to note here is that some people look at the second string of numbers and do not automatically notice that it conveniently groups into years. These people have just as much trouble with the second string as with the first – until the chunking is pointed out to them. This illustrates a number of interesting issues about working memory.

*Principle 7*: Working memory foothold ideas

    7.1   Working memory has a limited size, but it can work with chunks that can have considerable structure.

    7.2   Working memory does not function independently of long-term memory. The interpretation and understanding of items in working memory depend on their presence and on associations in long-term memory.

    7.3   The effective number of chunks a piece of information takes up in working memory depends on the individual's knowledge and mental state (*i.e.*, whether the knowledge has been activated).

This example also points out something of general importance. The study of the memorizability of number strings is an example of what I referred to above as a zero-friction experiment. It does not much resemble a real world activity but it tells something about fundamental brain mechanism. However, the zero-friction result can be trumped by friction when imbedded in a realistic situation. Since the three years in the second string-of-numbers example are tied to strongly held and easily recalled semantic knowledge, they are easily reconstructed at a later time. This is why coherence is so important in learning structured bodies of knowledge such as science.[*]

Item 7.2 in our list is fairly obvious when we think about reading. We see text in terms of words, not in terms of letters, and the meanings of those words must be in long-term storage. Item 7.3 is something we will encounter again and again in different contexts: How students respond to a piece of information presented to them depends both on what they know already and on their mental state – what information they are cued to access.

### 3.1.2    Compiled Knowledge: Creating Chunks

The number of chunks a piece of information has for an individual depends not only on whether or not they have relevant associations but how strong and easily that knowledge is activated in long-term memory. When a group of knowledge elements – facts and processes – is easily available and can easily be used as a single unit in working memory, we say the knowledge is *compiled*. Computer programming is a reasonably good metaphor for this. When code in a high-level computer language has to be translated line-by-line into machine instructions, the code runs slowly. If the code is compiled directly so that only machine-language instructions are presented, the code runs much more quickly.

Some of the difficulties students encounter – and that we encounter in understanding their difficulties – arise from the issue of compilation. Physics instructors work with many large blocks of compiled knowledge. Because of this, many arguments that seem simple to them go beyond the bounds of working memory for their students. If the students have not compiled the knowledge, an argument that the instructor can do in a few operations in working memory may require the student to carry out a long series of manipulations, putting some intermediate information out to temporary storage in order to carry out other parts of the reasoning.

### 3.1.3    Buffers: Holding Information in Working Memory

Studies with animals and with brain damaged amnesia victims make clear that working memory and long-term memory are carried by different structures in the brain.[†] Studies with subjects trying to recall strings of information indicate that items fade from working memory in a few seconds if the subject does not try to remember the information by repeating it consciously. [6] This working memory repetition is known as *rehearsal*. Think about looking up a telephone number in a phonebook. Most of us can't remember it – even for the few seconds needed to tap in the number – without actively repeating it.

The short lifetime of working memory has serious implications for the way we communicate with other people, both in speaking and writing. In computer science, holding information aside in preparation for using it later is called *buffering*, and the storage space in

---

[*]    Note this also gives us some predictability. Asked the string of years a week late, it is unlikely that an American would make the error "1756" for "1776" but might recall it as "1789" instead – substituting another data associated with the creation of the USA rather than making a random substitution.

[†]    From the response of patients and animals with damage to their brains, it seems clear that the structure that performs the translation from working to long-term memory lies in the medial temporal lobe, perhaps in the hippocampus or related structures.[30][24]

which the information is placed is called a *buffer*. Since human working memory is a buffer that is volatile and only has a lifetime of a few seconds, it can be very confusing to present people with information that relies on information that has not yet been provided. The given information may be forgotten by the time it is needed. Doing this can mess up a student's ability to make sense out of a lecture or a surfer's ability to understand a webpage.[*]

Learning something complex like how to solve physics problems involves more than just the automatic process of remembering a conversation held earlier in the afternoon or remembering experiences from a trip to Paris. What is critical in long-term memory for physics is not just having the memory, but being able to use it and make the associations that bring it into working memory. This is of particular importance for us in understanding our students, since we want their knowledge of physics to be functional: We want them to know when and how to use it as well as to recognize it. To understand these more complex issues we have to leave the simple cognitive- and neuro-science experiments and turn to the educational phenomenologists, though we still want to retain plausibility with respect to what is currently known about neuroscience and basic cognitive processes.

What is clear from many educational studies is that learning of complex abstract information is neither simple nor automatic. It requires substantial repetition and practice to compile knowledge and it requires carefully designed educational environments to lead to appropriate patterns of associations in long-term memory. In order to understand how this works, we need to know something about how information is structured in long-term memory.

### 3.2 Links, structures, and context dependence

The basic structure in memory activation is association. This principle is well established both from a neural base [2] and from cognitive studies. [6] Activating one knowledge element or resource may lead to the activation of other related resource elements. If you are asked to list the names or four animals and then to list four objects beginning with the letter "b," you are very likely to include a number of animals (such as "bear" or "beaver") in your list. Since neurons can exist in multiple levels of activation, we are not surprised to find that activating one item from long-term memory can make it easier to activate other items, even if they are not consciously activated. This is known as *priming*. It represents a first simple level or associational activation.[†]

Other associations flow quite naturally when a particular bit of knowledge is activated. Collins and Loftus describe this process with the term *spreading activation*. [34] Since the activation of some neurons can enhance or inhibit the activation of other neurons or strengthen or weaken other neuronal connections, we should not be surprised to find complex activation patterns and subtle responses.

---

[*] In the theory of communications, this leads to the "given-new principle" in conversation and writing. [31] [32]

[†] Priming can occur even when the stimuli doing the priming are presented very quickly and not consciously noted. [5]

Activation of one resource therefore tends to lead to the activation of a cluster of related resources. In the cognitive literature there is a large body of work on these issues and a wide variety of terms are used, often inconsistently from one research team to another. I will therefore introduce a small number of terms and define them carefully.

I refer to any set of related resources that are activated together in a particular situation as a *pattern of association*. Two terms are commonly used in the cognitive literature to describe particular patterns of association: schema and (mental) model. Both describe patterns of association that are robust and are activated reliably in a variety of circumstances. I follow the notation of D'Andrade and call such a pattern a *schema*[*] if it is a "bounded, distinct, unitary representation" that is not too large to hold in working memory. I call a pattern a *(mental) model* if it consists of "an interrelated set of elements which fit together to represent something. Typically one uses a model to reason with or calculate from by mentally manipulating the parts of the model in order to solve some problem." [20, p. 151][†] "Model" is the more inclusive term: a schema is a simple model.

In order to try to create a baseline from which we can build, I define the idea of context in the crudest possible fashion consistent with our principle *3*. In the mind of a particular individual, *context* is the state of activation of each of the neurons in the individual's brain at a particular instant. We then have the *context dependence principle*:

> *Principle 8:* The activation of a particular resource in response to a presented stimulus can depend not only on the stimulus but on the context – the activation pattern existing in the brain when the stimulus is presented.

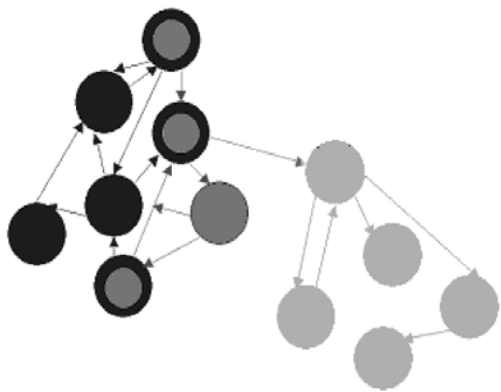

> *Fig. 4: A schematic representation of the association of resources into schemas. The different colored rings in the schema on the left indicates a context dependence. The darker circles associate as indicated under one context, the lighter circles under a different one.*

---

[*]  This term has a long history in the cognitive literature, going back to Kant [35] and Bartlett [36].
[†]  Note that Johnson-Laird uses the term "mental model" in a narrower, more technical sense. [37]

This principle is only a baseline and appears trivially obvious, especially since it contains the weasel words "can depend on." This doesn't say it does or it doesn't. This is part of what I mean by a "theoretical framework" as opposed to a theory. The principle tells us to pay attention to context dependence, but doesn't tell us how or when. That's the job of the various theories that need to be constructed. An illustration of a pattern of association relating two context dependent mental models is given in figure 4.

What the framework does it to tell us to ask of any theory: "What are you assuming about context dependence?"  And of course one must add, "and how do you show us that your assumptions are correct?" Theories of mind need to tell us what matters in the vast activation pattern and what doesn't for a particular activity in a particular situation. Although this is a weak principle, the focus on the individual and on mechanism does help us to keep in mind some obvious facts that are frequently ignored, such as:

> *Principle 9*: An item presented to a student is not part of the student's context until the student takes notice of it (though this "notice" may be through unconscious priming).

In other words, it does not suffice to consider the instructional environment and the material presented to the student to be the "context."  It is the student's response to the environment and what is presented that has to be considered to be the context for cognitive activity.

Since it is impossible to determine the complete activation context in an individual's brain, it is likely that we will often have to take a less deterministic view and to treat the activation of resources probabilistically.

The important thing to keep in mind is that the productive character of memory activation and the context dependence of the particular response lead to a system that has a general stability but a specific fluidity. The cognitive system – even for a single individual – is very labile (dynamic). It needs to be flexible when dealing with the immense complexity of the world. But we can also note (and we often tend to forget) that a lot of an individual's behavior is actually highly predictable. We just tend to take that part of behavior for granted. (For example: If you take your 8 year old child into a MacDonald's for the twentieth time, you are very likely to know exactly what she will want to order, how much of it she will eat, whether she will agitate to buy the "toy-of-the-week" and so on.)

Donald Norman described this well in the mid 1980s.

> *Because the schema is in reality the theorist's interpretation of the system configuration, and because the system configures itself differently according to the sum of all the numerous influences upon it, each new invocation of a schema may differ from the previous invocations. Thus, the system behaves as if there were prototypical schemas, but where the prototype is constructed anew for each occasion by combining past experiences with biases and activation levels resulting from the current experience and the context in which it occurs. [38, p. 535]*

I summarize this in the *productivity principle*.

> *Principle 10*: Resources and their organizational structures, schemas and models, are productively recreated at need out of activations of smaller elements of long-term memory.

### 3.3  Control and Executive function

The next component we want to identify from the cognitive/neuroscience literature is *executive function*. The associational patterns described above bring information from long-term memory into working memory in response to sensory data. But associations are often superficial and inappropriate. In addition, associations called up by sensory input can be contradictory. The world contains such a large number of possible objects, interrelationships, and options that our brain, large and complex as it is, could be easily overwhelmed by the number of choices we have to make, both of how we see and organize our view of the world and how we decide what to do in it.

#### 3.3.1     The Mental Executive and Selective Attention

Baddely conjectured that the brain has a *mental executive* that manages working memory. [6] The idea of executive function is very broadly supported by cognitive zero friction studies, lesion studies, and fMRI scans and appears to reside largely in the prefrontal cortex (PFC). [39][40] Executive function is in some sense a modern analysis of the concept of *selective attention*, which has been a topic of study in psychology since its earliest days. [41]

An example where you can feel this acting in your own head is the Stroop task. [42] If you are shown a series of colored blocks, it is easy to read off the colors. Your color receptors activate a response that is linked to an association with a word. If you are given a list of color words printed in black, it is easy to read off the colors. Your visual response activates semantic meaning associated with colors. If you are given the pattern shown in the top of figure 5 and are asked to read the color inks that the words are printed in you will find it easy since the word names and the ink colors are consistent. But if you try it in the lower figure, where they are inconsistent, you will find it extremely difficult.

In the latter case, your sensory input activates two contradictory color associations – one through your color perceptions, one through your word associations. The executive activity of selecting the activated response appropriate for the task is a function of the prefrontal cortex. Studies using fMRI show increased activity in the PFC during the Stroop task and patients with damage to this part of the brain have severe difficulty with the task. [43] Lesion studies also demonstrate that controlled social responses are also associated with the PFC. Patients with PFC damage often show reduced inhibitions and inappropriate social behavior.

#### 3.3.2     Controlling Activation of Associations

As in the study of associational patterns, there are a variety of attempts to build fundamental neuro-cognitive theories of executive function. [44][45] Again, we are not at this time interested in building a detailed theory from first principles, but rather in establishing a theoretical framework with some mesoscopic elements relevant for education. All executive functions have a structural feature in common. They enhance (turn on) or suppress (turn off) associational patterns.

As a result, we abstract from the studies of attention and executive function a fundamental structure: that of *control*. We can think of the analogies of two kinds of switches: a flashlight and a train on a track. In a flashlight, the battery and the bulb are associated by a link. The switch activates and deactivates that link. In a train on a track, the train may have an option of traveling in one of two directions: the switch chooses which way it will go. Of course, these are only analogies. Neurons can be in a variety of activation states and resources are active. As in our study of associational structures, this means that elements of executive control are potentially as productive and labile as schemas.

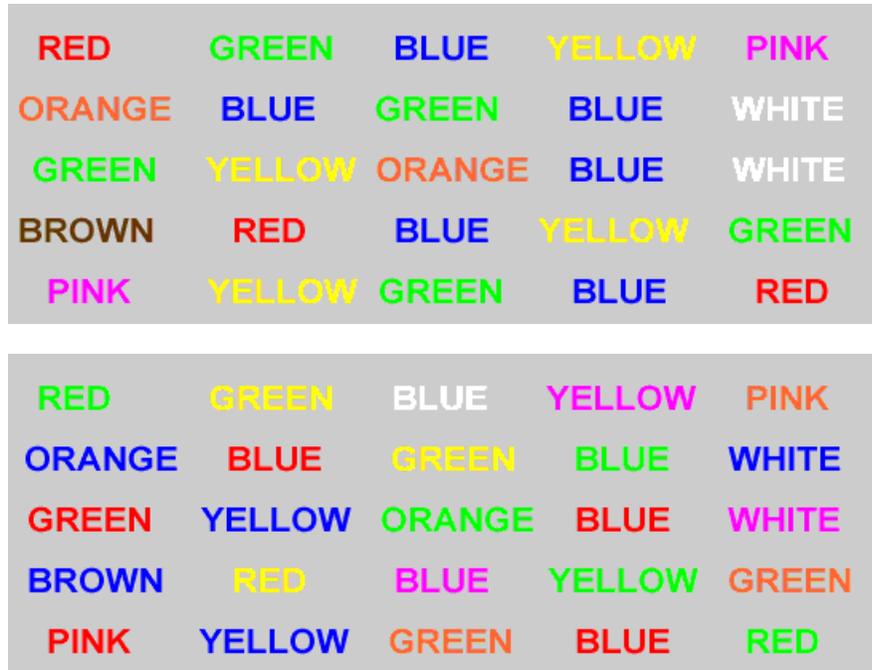

*Fig. 5: The Stroop task illustrating executive function. Read the colors that the words are printed in. In the top figure word and ink color are consistent. In the bottom they are not.*

Separating control from association allows us to describe knowledge structures and their context dependence separately and has powerful implications both for our understanding of our students' responses and for our understanding of what produces those responses.

The two basic structures in our cognitive model are illustrated schematically in figure 6. Association of resources provides the structure of knowledge appropriate to a given situation, while the control structure is associated with attention, context dependence, and goal-oriented decisions. This coarse separation provides the fundamental basis of our theoretical framework. Note that we are not here referring to neural structures. Both associational and control structures occur at all levels of brain functioning, down to the level of individual neurons. The separation into association and control is a $0^{th}$ order approximation for organizing our mesoscopic analysis of cognition.

These elements provide a basic structure for organizing what we know. In order to provide useful structures for understanding teaching and learning, each sector of our theoretical framework now needs to be fleshed out by specific models. In the next two sections we discuss some plausible starting points.

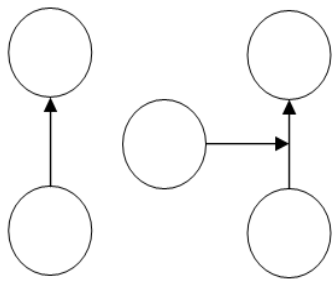

*Fig. 6: The two basic structures of our cognitive model: association (left) and control (right).*

## 4 Knowledge and Knowledge Structure

At its core, my theoretical framework describes student knowledge as comprised of cognitive resources in various forms and levels of hierarchy. Within each level is a collection of resources that are primed, activated, and deactivated depending on context and executive control.

*Principle 11*: Reasoning about any particular question entails a selection, tacit or explicit, from a collection of resources.

More details require a model of what the resources are and how they are associated. Various proposals have been made that assume that the relevant structures are tightly associated, loosely associated, compiled, etc. [46] [47] [48] [49] [50] In this section, I describe some elements of a resources model of knowledge structure that is appropriate for looking at students' knowledge of physics.

### 4.1 Identifying relevant resources: Reasoning and phenomenological primitives

In order to set up a model within our theoretical framework, we need to specify the resources that are available for the tasks of interest and how they are associated in relevant contexts. One model that has been proposed identifies a particular set of schemas as a set of resources relevant for physics learning: the *phenomenological primitives* (*p-prim* for short) described by diSessa. [48] These are basic statements about the functioning of the physical world that a student considers obvious and irreducible.

Asked to explain why it is hotter in the summer than in the winter, many students will respond that it is because the earth is closer to the sun. Educators often attribute this response to a faulty conception students have formed, by which the earth moves in a highly eccentric ellipse around the sun, and for some students this may be the case. DiSessa's account allows an alternative interpretation: Asked the question, students conduct a quick search among the resources they have in their knowledge that may apply, and one of the first they tend to find is the notion that *closer is stronger*. Students' tendency to explain seasons in terms of proximity to the sun may be understood as a faulty activation of this resource, which in itself is neither correct nor incorrect. (It *is* hotter closer to a fire.)  Other resources students have available would be more productive, such as for understanding greater strength arising from more direct incidence. DiSessa has described some of these primitives, and there are certainly many more.

These structures are "primitive" in the sense that they are indivisible to the user: Ask a student why it feels hotter closer to the fire, and from the student's perspective there is nothing else to say.[*] That's just the way things are. From this perspective, there are thousands of primitives – ones for interaction with the large variety of physical experiences we have in this complex world. DiSessa explicitly [48] chooses not to separate abstract and concrete primitives, but to focus on irreducibility as viewed by the user.

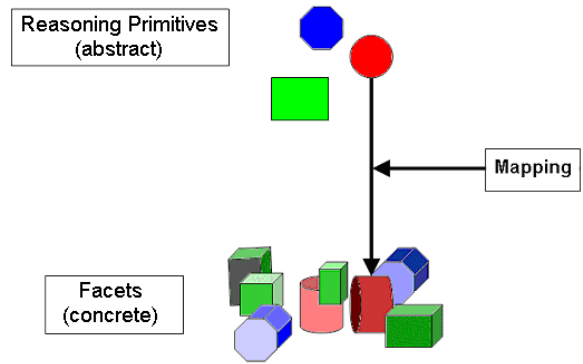

*Fig. 7: Abstract reasoning primitives are mapped into facets for specific physical situations..*

Having a large number of primitive elements is somewhat daunting in what is meant to be a theory of mechanism. I therefore find it convenient to abstract *reasoning primitives* from diSessa'a p-prims. These are general rules and relationships that become concrete statements about a particular physical system as a result of a *mapping* – an association of the elements in the reasoning primitive with an item or quantity in the physical world. (See figure 7.) Following Minstrell, if I am focusing on the result of the mapping I refer to the mapped

_______________________

[*]      A physicist who has well-developed ideas about the mechanism of heat generation and transmission m Although it is difficult to obtain direct evidence for a reasoning primitive and a mapping when the structure is invisible to the user, it is nonetheless useful as a classification scheme. In addition, it has the satisfying character that there are a fairly small number of reasoning primitives. It is the mapping onto the complexity of the world that produces the huge number of facets.

ay be able to construct reasons for the result. It is no longer a primitive result for her.

primitive as a *facet*. [50] Thus, in the summer/winter example the reasoning primitive is *closer to a source is more effective*. The result *when it is warmer on the earth we are closer to the sun* is a facet. When I want to emphasize the irreducibility to the user, I continue to refer to the principle as a p-prim.

Students are rarely aware of the structure of a mapping. An unusual but felicitous counterexample to this rule is provided in the transcript of an interview done by Loverude probing student understanding of the principles of fluid mechanics. [51] Loverude posed the problem on the flow of water from one tank to another shown in figure 8. The student's response dramatically illustrates the idea of reasoning primitives and mapping.

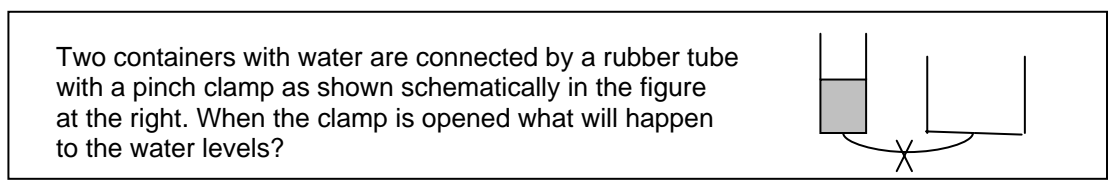

Two containers with water are connected by a rubber tube with a pinch clamp as shown schematically in the figure at the right. When the clamp is opened what will happen to the water levels?

*Fig. 8: A problem is fluid mechanics posed by Loverude. [51]*

*It will continue to move until there's some sort of homeostasis, or whatever you call it, an equilibrium, which may be, it's either going to be that the volumes are the same, or the heights will be the same, I'm trying to think of which one it's going to be. Do you want me to tell you which? (laughs) Um, I'll go with volume for now but I'm not sure. So it'd be lower in the larger container....*

*It's really not reasoning, it's more just thinking about, these things want to be equal, and what are the ways things could be equal? One was fluid volume, and one was height.*

This student explicitly is activating *balancing* and is explicitly (and unusually) aware of the need to map the abstract primitive onto the quantities in the problem. Although this is not common, it is a nice confirmation of what we believe is happening in general at a less conscious level.

### 4.2 Associational Patterns: Theories of student knowledge structures

Once we've identified relevant resources, the next natural question in our theoretical framework is: to what extent and under what circumstances (contexts) are resources associated with each other. Does activation of one resource imply the activation of others in most cases? If the patterns are tightly organized, so that activation of one element almost always activates the others, we may treat the pattern as a schema. (Recall that we use the term *schema* as a general term to refer to a robust pattern of association of knowledge structures.) If the pattern is sufficiently tightly organized (compiled), it can serve as a single unit in working memory. An issue of great interest in physics education research is the extent to which student knowledge resources are tightly and robustly organized and how strongly they are activated by a range of situations.

In the early 1980s, education researchers proposed that students' knowledge of physics (especially in mechanics, where everyone has everyday experience with forces and motion) were organized into naïve theories that were robust and coherent. These theories often were assumed to resemble the theories constructed by natural philosophers throughout history. McCloskey and Vosniadou are two proponents of this viewpoint. [52][53] This theory is often referred to as the *misconception*, *naïve conception*, or *alternative conception* theory.

An alternative hypothesis is that student knowledge and reasoning consists of weakly organized resources. In this case, student reasoning might appear fragmented and inconsistent. McDermott and diSessa are two leading researchers who have argued strongly that empirical evidence supports this theory. [48][54]

Within our theoretical framework, it is clear that the characterization of a well organized knowledge structure as a "misconception" (or even a naïve or alternative conception) focuses on an issue that is not part of the cognitive model – whether or not the knowledge structure agrees with the one we are trying to teach. It is more appropriate to characterize the knowledge along the axes of <u>robustness</u> (how broadly the knowledge is activated in a variety of situations), <u>degree of compilation</u> (the extent to which complex knowledge can be applied as a unit in working memory), and <u>level of integration</u> (how much diverse knowledge is tied together). Rather than characterizing the two theories described above as "misconceptions" and "fragmented," we will describe them as the *model theory* and the *modular theory*. The idea is that in the first theory, student resources are assumed to be well organized into a stable cognitive model of how a system works, while in the second, knowledge is more fragmented and labile. Some characteristics of these two theories are outlined in the table below.

| MODEL THEORY OF STUDENT KNOWLEDGE | MODULAR THEORY OF STUDENT KNOWLEDGE |
|---|---|
| Strongly associated resources include many elements | Activated resources are largely independent and only weakly associated |
| Activated in a wide variety of contexts | Activated in limited contexts |
| Stable and resistant to change | Labile and easily changed |

*Table 1: Two contrasting theories of student knowledge structures. Adapted from [55].*

Notice that these two theories are not the only possibilities. For example, there could be a single resource that is activated in a wide variety of contexts and is stable and resistant to change. I propose that the term *misconception* be reserved to mean a knowledge structure that is activated in a wide variety of contexts, is stable and resistant to change, and is in disagreement with accepted scientific knowledge.

Notice also that the model theory of student knowledge is much closer to the kind of knowledge we want them to develop than the modular theory is. If students start with an expectation that they will develop a coherent understanding we can instruct them in a very different way than if that is not a part of their way of thinking about the world. I say more about this in section 5.

An example of what I mean by "labile" is seen in a transcript of a laboratory activity observed by my research group. The lab takes place in the second semester of college physics. The two lab partners are upper division biology majors carrying out a traditional laboratory on the interference of light. The TA has asked them to explain the behavior of the pattern produced on a screen by the interference of light from a laser as it passes through two narrow slits as the separation between the slits is changed.*

> *Veronica: It [the pattern] gets more, well, obviously when it gets more narrow, the slits get narrower. As it gets wider, the slits get wider. Wow. [sarcasm]*
> *Claude: The width of the slit increases with increasing? So, what's that?*
> *Veronica: This is .2, the other one was .4 Or .02*
> *Claude: That's .02?*
> *Veronica: Yeah.*
> *Claude: Then [the pattern we got with] 8 was narrower. Yeah. We started out with .08 and it was like…*
> *Veronica: Oh, the <u>wider</u> the width the narrower the slits get. Because there's more room for light interference because there's more rays of light going through.*

Veronica's first response appears to be a facet mapped from the reasoning primitive "more cause produces more effect." Notice how quickly and easily Veronica switches from being very confident that the dependence works in one direction to being very comfortable and confident that it works in the opposite way. She is able to easily re-map her reasoning primitive to obtain the opposite result.

More complex situations can also be observed. A very nice example comes from the work of Scherr and collaborators. [56][57] In this work, the researchers probed the knowledge of upper division physics majors and graduate students who were studying special relativity on the topic of simultaneity. They found that most of the students believed

1.  Events are simultaneous if an observer receives signals from the events at the same instant.
2.  Simultaneity is absolute. If one observer sees two events as simultaneous all observers will see events as simultaneous.
3.  Every observer constitutes a distinct local reference frame.

In interviews, they asked students to consider the problem shown in figure 9.

In discussing this problem, students clearly demonstrated resources 1 and 3. They treated the individual as a local reference frame, failing to remove the effect of signal travel time and failing to consider the global reference frame of rods and clocks that had been taught in lecture. However, when the question was reformulated to specifically mention light travel time and students probed to recall the technical meaning of the term "reference frame," they had no trouble correcting themselves.

---

* Throughout this paper, names cited in transcripts are gender correct pseudonyms.

Mt. Rainier and Mt. Hood erupt at the same time in the reference frame of a seismologist at rest in a laboratory midway between them. A spacecraft flying past Rainier towards Hood at $v$=0.8$c$ is directly over Mt. Rainier when it erupts. Let Event 1 be "Mt. Rainier erupts," and Event 2 be "Mt. Hood erupts."
In the spacecraft frame, does Event 1 occur before, after, or at the same time as Event 2?

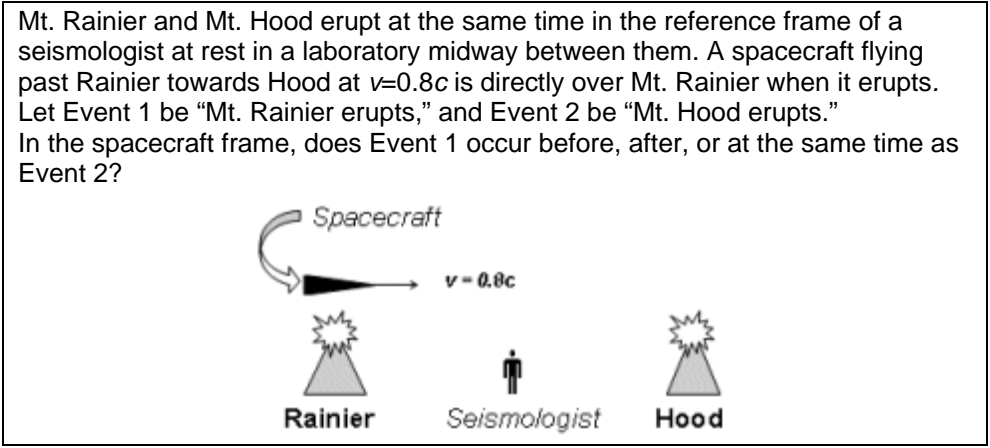

*Fig. 9: Problem posed by Scherr et al. [56]*

The second student belief tells a different story. Scherr et al. posed the problem shown in figure 10 to their students working in a Tutorial environment.* This is the standard Einstein railroad car simultaneity problem with a twist. Students who believe item 2 don't appear to be troubled by getting different results for the wavefronts. (Perhaps they don't interpret wavefronts to be tangible objects.) But the tape recorder is a different story. Here is a typical transcript of an exchange where four students are considering this problem.

> *Dennis: We just figured out that the tape player plays in Alan's frame.*
> *Tony: But it can't. In Beth's frame they hit her at the same time. So she won't hear it.*
> *Jana: But look down here, it's asking if she hears it and if the tape will have wound from its starting position. If the tape is going to play, that's it, it's going to play.*
> *Tony: But it can't play for Beth! She's in the middle! They hit her at the same time!*
> *Dennis: But we just figured out that it plays!*
> *Tony: Right! And then a black hole opens up! And God steps out! and he points his finger and says [shouting] "YOU CAN'T DO THAT!"*

On the issue of the absoluteness of simultaneity, student responses are highly stable and resistant to change. In interview situations, Scherr et al. found that students sometimes sit silent and unresponsive for thirty seconds or more when considering this problem. In groups, they observed students sometimes becoming loud and aggressive, or resorting to absurdities (as above) or to topics they know are confusing and that they may not understand very well (such as quantum mechanics).

---

* The University of Washington Tutorial environment is a set of worksheet-guided small-group interactive-engagement lessons that often make use of cognitive conflict. For a description see [58] or [88]. A collection of these lessons are published in [64].

A train is moving along a long straight track at a high velocity (near the speed of light). Alan stands on the ground by the train track; Beth stands on the train. Sparks jump at front and rear of train, simultaneously in Alan's frame, leaving char marks.

       -- Draw the light wavefronts produced by the sparks in Alan's frame.

       -- In Alan's frame, do the wavefronts hit Beth at the same instant?

       -- Does Beth see the wavefronts arrive at the same instant?

A cassette player sits at Beth's feet. If the front wavefront hits it first, it plays Beethoven's Fifth Symphony at top volume; if the wavefronts hit simultaneously, it does not play.

       -- Does the player play in Alan's frame?

       -- In *Beth's* frame?

Later in the day, Beth ejects the tape from the player. She descends from the train, and she and Alan examine the tape together.

       -- Will the tape have wound at all from its starting position?

*Fig. 10: Relativity problem posed by Scherr et al. [56]*

This example illustrates that even on a reasonably narrow instructional topic, we may encounter student schemas that are modular (weakly held, fragmented, labile) at almost the same time that we activate misconceptions (strongly held, robustly activated, stable, and wrong). Our model can accommodate both types of structure. The decision as to which is more appropriate is an empirical one – and could be different for different topics, individuals, or situations.

*4.3 Linkage patterns: Building coherence and integrating knowledge*

There have been a few attempts to describe the way students associate their primitive resources. Often, however, terms such as "concepts" are tossed around without definition assuming that "everyone knows what it means." This can lead to confusion and substantial wasting of time arguing when two researchers use contradictory tacit definitions.

In an attempt to begin to define appropriate associational structures for use in understanding physics learning, diSessa and Sherin have introduced the concept of *coordination class*. [49] In [59], diSessa defines a coordination class as "a specific collection of knowledge and strategies that allow us to read out ('see') a distinctive class of information from the world." DiSessa and Sherin structure a coordination class into two parts in order to permit the description of the context dependence of an idea. A *readout strategy* is a set of resources that translate sensory information into meaningful and processable terms. A *causal net* is the set of relevant inferences about the relevant information and their context-dependent associations. Figure 4 illustrates some elements of what might be the causal net for a coordination class.

One example of a coordination class is 'force.' DiSessa discusses this in [59] and I refer you there for details. Wittmann applied the idea of coordination class to his description of student understanding of waves on a string. [60] At this summer school, Mestre reported on an application of the coordination class structure to students' understanding of the ball-on-a-track problem; I discuss this in detail in section 5.2.2 (see figure 15). [61] In discussions with attendees to this summer school in response to Mestre's presentation, it was clear that the issue of whether the community wants to agree to define a coordination class as a stable or dynamic (labile) structure has not been decided at this time. My sense of diSessa and Sherin's intent in introducing the term is that they were particularly concerned to be able to describe the context dependence of phenomena. In that case, it appears to me to be more productive to define a coordination class as a labile, dynamic structure that has associational links that could be weak or strong. As, for example, a student made the transition from a naïve user of the term 'force' to an expert user one would say that his coordination class evolved rather than that he moved from not having a coordination class to having one. For more details, see [59] and [61].

The coordination class structure, especially if one accepts a dynamic definition, is particularly appropriate for discussing the transition from modular to model-based reasoning. Another kind of associational structure has been proposed by other researchers who assume that student reasoning with mental models is dominant. Samarpungavan and Wiers introduce the idea of an *explanatory model*.[*] [62] This consists of a network of prestored "beliefs" (possibly interpretable as facets in my framework) that constrain the particular associational patterns a student can activate. The authors assume that such a framework is internally consistent and robust. Although it takes a bit of translation to get this model into my theoretical framework, I would interpret their description as saying that the fundamental resources that make up explanatory models are stored directly without being generated on the spot by the mapping of a more abstract resource.

An example of how associational patterns affect the performance of students in physics classes is given by Sabella. [63] The problem shown in figure 11 is most conveniently solved using the work-energy theorem. Sabella found in interviews that some graduate students doing this problem got 'stuck' in a pattern of associations to forces and could not activate their knowledge resources about energy. Parts a) and b) in the problem seem to play a strong role in activating force resources (and perhaps suppressing energy resources). Introductory students in calculus-based physics did significantly better on this problem (and used energy arguments much more frequently) when parts a) and b) were omitted. [63]

---

[*] The authors use the term "explanatory framework." Since I am trying to use the terms "framework," "theory," and "model" in a consistent way, I have translated this to my terminology. Their explanatory framework is a mental model in my terms.

---

A hand applies a force to a small 1 kg block from "A" to "C." The block starts at rest at point "A" and then comes to rest at point "C." The block moves along a frictionless surface from "A" to "B" and then travels an <u>equal</u> distance along a surface with friction from "B" to "C" with the force of the hand <u>remaining constant</u>. The force of the hand is 2 N to the right and the distance from "A" to "C" is 2 m.

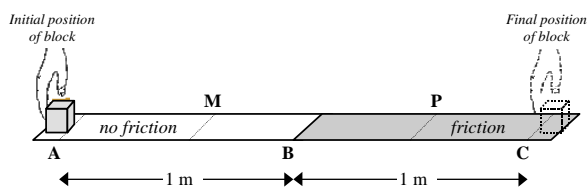

a. Draw a free body diagram for the block when it is at "P."
b. Is the magnitude of the net force acting on the block at "M" greater than, less than, or equal to the magnitude of the net force acting at "P"? Explain your reasoning.
c. i.) Draw a vector representing the acceleration of the block at "P."
      If the acceleration is zero state that explicitly.
   ii) Does the magnitude of the acceleration increase, decrease,
      or remain the same as the block moves from "B" to "C"? Explain.
d. Calculate the coefficient of kinetic friction μ.

---

*Fig. 11: Problem to study student associational patterns of force and energy. [63]*

Whether or not one accepts any particular associational structure to describe a particular knowledge set, what is essential is the central and well-established insight that reasoning about any particular question entails a selection, tacit or explicit, from a collection of resources. All of these resources are useful in some contexts, or they would not exist as resources. Reasoning involves selecting those that are useful in the given situation. Learning physics, in this view, largely entails a reorganization of existing resources. For this reason, it is critical that educators attend to how students select from, coordinate, combine, and transform their cognitive resources. This is the second component of our theory.

## 5    Epistemology and Expectations

My second level of cognitive structure is control. The world we live in is too complex to attend to every item in it (whatever an "item" is). Presented with some environmental stimuli, individuals respond by parsing the world into pieces, paying attention to some parts and ignoring the rest. This parsing is more than just deciding what to look at: it includes deciding what behavior is appropriate in a given situation. There is something "between" our coding of sensory input data and our making sense of that data – a control filter that decides what knowledge structures are to be activated. Sometimes these decisions are explicitly conscious and sometimes they are automatic and unnoticed. This control structure has strong overlaps with the traditional ideas of epistemology and expectations but it divides somewhat differently than is traditional.[*]

Since we are concerned with teaching and learning, students' ideas about knowledge and knowledge construction – what is often referred to as *epistemology* – are of great importance to us. Users of the term "epistemology" often fail to distinguish *declarative epistemology* – knowledge structures that are statements about the nature of knowledge – from *functional epistemology* – control structures that choose how an individual will construct knowledge in a particular situation. The failure to make this distinction can lead to substantial confusion about the meaning of the term "epistemology."

Most instructors are aware that student expectations about what they are supposed to do in class can play a powerful and destructive role in creating "selective attention" that limits what students actually do in class and what they learn. These expectations interact strongly with (and often control) the resources the students have for creating knowledge.

The cognitive level of control structures provides an appropriate theoretical framework for developing models of these concepts. The Maryland Physics Education Group has been developing a model of control elements relevant for education based on work done in educational psychology [65], sociology [66] [67], sociolinguistics [68], and discourse analysis.[†] [69] Since we are interested in teaching and learning, we will focus on resources and control structures associated with building new knowledge.

Most research on students' understanding of the nature of knowledge creation in science is about what students articulate about the nature of knowledge and learning. There are two hidden assumption in much of this work. (1) These articulated statements are fixed, robust, and often wrong: "misconceptions" of epistemology.[‡] (2) These articulated statements affect students' choice[§] of knowledge building tools. Our theoretical framework alerts us to the idea that these assumptions may not be correct. For the first, there may be considerable context dependence, even in expressed, articulate statements that students make about knowledge and knowledge construction. For the second, articulated beliefs may play little or no role in the activities in which students choose to participate in order to construct scientific knowledge for themselves. The first statement describes declarative knowledge, the second a control activity – the selection by whatever means (conscious or implicit) to activate particular knowledge structures. These two hidden assumptions conflate declarative and functional epistemology.

---

[*]   In any case, these terms are often used in the literature inconsistently.

[†]   In this context, "discourse analysis" refers to the analysis of text and how it is interpreted rather than to the analysis of conversational interactions.

[‡]   Such "epistemological conceptions" are often referred to as *beliefs*.

[§]   Recall that "choice" here and elsewhere in this section refers to neuronal control structures and is not intended to imply a conscious action on the part of the student.

We need a finer grained analysis of the resources associated with knowledge construction and with the control and activation of these resources. There are three components to our theory: epistemic resources, frames, and messages. The first idea we need is that there are resources for knowledge building. Following Collins and Ferguson [65], we refer to these epistemic resources as *epistemic games and forms*. The second idea we need is that there are control structures that manage the activation of associations of these resources. We refer to the relevant structure for choosing knowledge building tools as *epistemological framing*.

Although framing is a process in the cognition of individuals, it is the individual's response to input from the external world, it depends on input from the physical world, from culture, and from social interactions. We therefore need a third idea to analyze the interaction between the individual and the socio-cultural environment –communication paths that facilitate these interactions. The critical concepts here are the *messages and metamessages* (messages about how to interpret messages) that individuals receive from interactions with the world. These three ideas, epistemic resources, epistemological framing, and messages, combine to permit a very rich and revealing analysis of student knowledge construction. They provide the tools to carry out an "epistemological discourse analysis" – a parsing of the process by which students construct knowledge – that allows us to describe and understand both what is happening in an individual student and in an interaction between students.

*5.1 Epistemic Resources*

Individuals have a wide variety of resources for constructing knowledge. A small child may know what's for dinner because "Mommy told me" (*knowledge as propagated stuff*). She may know her doll's name because "I made it up" (*knowledge as fabricated stuff*). [70] A student may "know" that a big car hitting a small car exerts a bigger force on the small car than the small car exerts on the big one because "the big one is stronger" (*knowledge by p-prim*). We refer to the processes and tools students use to decide they know something and to create knowledge as *epistemic resources*.

Our theoretical framework reminds us that with any discussion of knowledge resources, an understanding of associational patterns (schemas and models) of those resources are also useful. Two such structures (introduced by Collins and Ferguson [65]) are epistemic games and forms. An *epistemic game* is a coherent activity that uses particular kinds of knowledge and the processes associated with that knowledge to create knowledge or solve a problem. Thus a physicist may play a "making meaning with mathematics" game to map a physical situation onto a set of equations, manipulate those equations to obtain a solution, take limiting cases to check against his intuition, etc.

An *epistemic form* is an external structure or representation and the cognitive tools to manipulate and interpret that structure. In our complex and technical society, much of the information building we do uses a variety of external representations for knowledge construction. An abacus, slide rule, free-body diagram, and a position-velocity graph are all epistemic forms. When I refer to an epistemic form, I really mean the cognitive structures that we possess for using and interpreting the results of manipulating the structures, but I will sometimes talk as if the form is the external structure itself.[*]

Collins and Ferguson limit their use of the terms epistemic games and forms to something used by experts that we are trying to teach our students to use. We extend the term to include activities and tools possessed by or created by students spontaneously. Our use of the terms is therefore descriptive as opposed to normative. As an example, consider Veronica's use of p-prims to answer the question "explain the behavior of the interference pattern as the separation of the slits is varied" that is presented in the transcript in section 4. There I focus on the lability of her use of p-prims. Here, I focus on her choice of p-prim reasoning to answer the question. The use of p-prims to answer an "explain" question is an epistemic resource that I call *p-primming*. The knowledge and the associated decision to choose to use p-primming in a particular problem situation is epistemological. (I will refer to the process of choosing to use related p-prims to reason with as *making common sense*. This is an e-frame as described in the next section.)

As with any knowledge resources, principle 8 in section 3.2 suggests that epistemic resources should not be thought of as always available but that they are likely to be activated in a context-dependent way. This suggests that we ought to focus on functional rather than on declarative epistemology.

An example of the fluidity of the activation of epistemic resources is given by Hammer in his paper in this volume. [71] The incident is observed in an activity in an American 8[th] grade classroom. The class had been studying geology and the teacher had organized the students into groups of eight to prepare a poster describing the rock cycle.[†] The video data that Hammer describes captures the students seeking information in their notes and text, struggling with complex words they do not understand. (One student repeatedly refers to "Teutonic plates.") They don't seem to be making much progress and few students are involved. The teacher suggests that they might try to use their own knowledge and understanding, rather than trying to look things up. The result is a dramatic shift in the students' behavior. They start describing what they know in their own words and become increasingly engaged. Some students who had previously sat silently now fully enter the exchange. I would describe this sequence of events by saying that the students shifted the epistemic resources that they had activated for solving the problem from *by authority* to *sense making*.

One of the most important components for our understanding of learning is the epistemological component:  the student's expectation as to what epistemic resources are appropriate to use in an instructional environment. To be able to describe these choices and shifts, we need an additional concept: framing.

---

[*]   In linguistics, this is referred to as *metonymy* – letting an associated object stand for something more complex; like referring to "the White House" when we mean the American presidency.

[†]   A topic in earth science describing the creation of rock from magma through volcanism and the absorption of rock to magma through plate subduction.

*5.2  Epistemological Frames and Framing*

An individual's expectations activate what she pays attention to and what she ignores in response to the 10,000 things and their interactions – a selective filter. A college student has had many years of schooling and (thinks) she knows what to expect when she walks into a classroom. If the students' expectations about what they are supposed to do fail to match the teacher's expectations about what the students should do, both may be disappointed. In many of the papers in this volume, we find examples in which students have the resources to answer physics problems but do not activate them appropriately – and may continue not to do so even after it is suggested to them. To understand student expectations and the role they play in learning, we introduce the control structure of "framing."

The idea of *frames* and *framing* was proposed by psychoanalyst Gregory Bateson [66] and anthropologist Irving Goffman [67] and used by socio-linguist Deborah Tannen [68] to describe how individuals develop expectations that help them make sense of complex world situations, especially social ones. According to Goffman, a *frame* is the individual's answer to the question "What is it that's going on here?"  Bateson first introduced the idea in the context of observing sub-adult monkeys scuffling and nipping at each other. They appeared to know the difference when a particular behavior was to be interpreted as play and when the same behavior should be interpreted as fighting or aggression. He called the monkey's interpretation of the situation a *frame*.[*] Tannen describes a pediatrician discussing a child's illness with the child's mother in the presence of the child and a tape recorder for the doctor's permanent record. The doctor's interaction with the three (child, mother, and tape recorder) were each carried out in different frames, using different vocabularies, levels of sentence complexity, and tones of voice.

An example of how this works is in what we would call in our current framework, *the restaurant frames.* [72] When you enter a restaurant, you have a variety of resources on which you can call. You know how to read a menu, either one printed on paper or displayed on the wall (and perhaps you know how to recognize a "daily specials" menu chalked on a slate). You know how to eat with your hands or with a knife and fork (and perhaps with chopsticks). You know how to pay for your meal with cash or a credit card. When you enter a particular restaurant, a variety of sensory impressions lead you to automatically activate a subset of these resources together with an expectation that you will carry them out in a particular order. Two examples are shown in figure 12. At the left is shown the behavioral model activated by entering a fast-food restaurant; at the right, the model activated by entering a formal restaurant. Most adults living in cultures that have restaurants of the type shown in the examples in figure 12 have all the resources the need to function in either type of establishment: but they only activate the appropriate ones in each case.

---

[*]  This observation was important at the time, since American psychology was in thrall to a stimulus-response model of behavior (behaviorism). Many psychologists considered that unobservable mental states should not be considered. Bateson's example was an reasonably clear case where the same stimulus led to different responses depending on the animal's internal mental state.

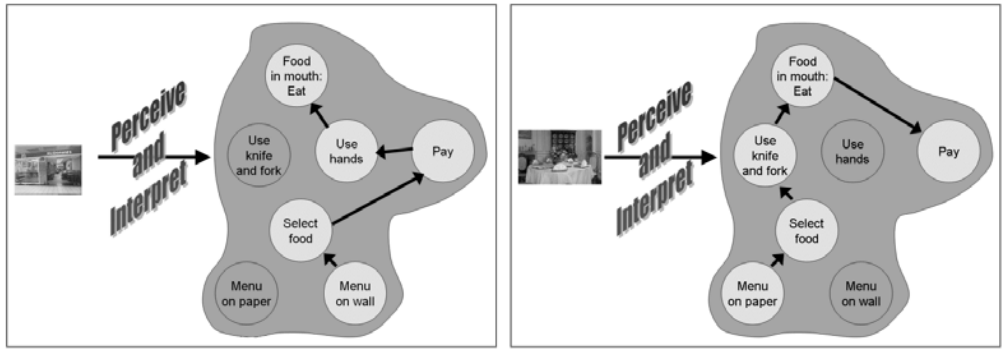

*Fig. 12: The behavioral models activated by entering two different restaurants.*
*Left: a fast food restaurant; right, a formal restaurant.*

In either case, long-term memory elements are being activated. I refer to the <u>process</u> of perceiving, interpreting, and activating a particular set of long-term memories for dealing with a situation as *framing*.

Frames describe an individual's expectations: they are the way individuals select, organize, and respond to the situations in which they find themselves. Framing is part of an individual's cognition <u>but:</u>

- Framing is carried out by the individual in response to social and physical experience.
- Framing responds strongly to (sometimes subtle) social cues.
- When interacting individuals frame a situation differently, it can cause serious communication problems.

When students enter a classroom, they frame what is going on. This framing process has many components:

- a social component (Who will I interact with and how?),
- a physical component (What materials will I be using?),
- a skills component (What will I actually be doing?),
- an affect component (How will I feel about what I'm going to be doing?), and
- an epistemological component (How will I learn / build new knowledge here? and What counts as knowledge here?).

Since our primary concern in studying education is understanding teaching and learning, the epistemological component of a student's frame is particularly important to us. Let's consider some examples of students' epistemological framings and how they can cause difficulty in the classroom.

A student in an inquiry-style Tutorial [64] or laboratory [73] might be asked to "make a prediction." I have seen students in this situation pull out notes and textbooks in order to find the answer when what the instructor wanted and expected was for the student to think about the situation and use their best intuition to create a prediction. Student and teacher had framed the situation differently, with the result that the student activated a different epistemic resource than the teacher intended (*knowledge from authority* vs. *knowledge as fabricated stuff*).

We define an *epistemological frame* or *e-frame* as the set of epistemic resources the individual assumes is appropriate to carry out the task at hand. Note that we have separated *epistemic* (describing the act of creating knowledge or the decision that something is known using various processes) from *epistemological* (describing statements about how something is known or the decision to use particular knowledge construction tools).

Note also that e-frames can be fluid, evolving to the situation quickly and easily in the way that schemas of knowledge resources may change and adapt. In other cases, they can be quite rigid. The question as to whether an e-frame is fluid or rigid is analogous to the question as to whether some associational knowledge pattern is modular or a model. Which it is needs to be determined empirically in particular situations. The potential (and common) fluidity of e-frames implies that a student might e-frame different parts of the class differently.

*5.2.1    Example: A student's e-frame makes a difference in how they learn*

Paying attention to epistemic resources and epistemological frames helps us be aware that a student's difficulty with a particular problem or activity might not be associated with the student's knowledge but with the student's framing of the situation. We saw an example of this at Maryland when we observed a group of students working together on a UW Tutorial [64, p. 155]. Tutorials are group activities in which students work together in a guided fashion (with worksheets) to build qualitative reasoning skills. In the transcript below, a group of four students are working through an activity in which the students are being guided to build a mental model of the propagation of light in straight lines. This lesson occurred about half-way into the second semester. One activity has the students holding a small bulb in front of a mask (a black piece of cardboard with a small hole in it) and trying to predict what pattern of light and shadow would appear on the screen (a white piece of cardboard) and how the pattern would change when the bulb was moved up. The apparatus is sketched in figure 13. Veronica, Jan, and Claude had not had my class first semester, but they had already had a half-dozen tutorial lessons of this type.

> *Claire: But what's the normal direction of the light? Cause that's what I'm asking.*
> *Veronica: It spans out, and whatever passes through that circle is the part we're going to see.*
> *Jan: So the light is like that [drawing], and these are the rays, and the vector that points that way is going to go through the hole.*
> *Claire: Ok, so then if you move it up then it's going to be…?*
> *Claude: … [unintelligible]…the light…*
> *Jan: Right. So like it has…[pause]*
> *Veronica: Really, it's just normal.*
> *Jan: All the rays are going like this. So, it's kind of like polarized.*

*Veronica: [pause] Mmmm. Not really. [long pause] It's just, well, it's just, guys, you're making it, you're trying to make it too difficult. It's just, the light goes out. It only goes through that one circle. So, obviously, if it's down here [pointing to the screen], and I'm looking [back towards the light] through that circle… Look. You're sitting down here [pointing to the screen]. You're looking up through that little circle [pointing to the mask]. All you're going to see is what's up there. It's a direct line.*
*Jan: [overlapping] Look, I see what you're saying, right? But I'm just trying to make it like physics – physics-oriented. [laughs]*
*Veronica: [decisively] It <u>is</u> physics-oriented. That's just the way it is.*
*Jan: [in a low voice] Okay.*

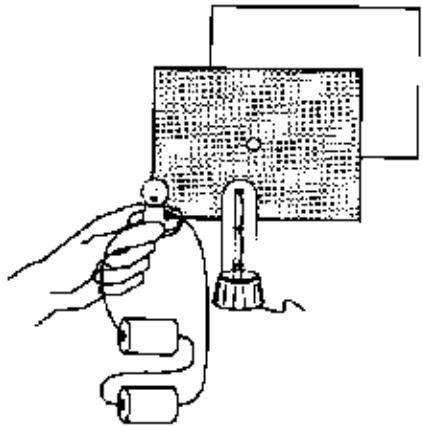

*Fig. 13: Apparatus for a Tutorial in learning to reason about light and shadow. [64]*

In this transcript, one interpretation of what's going on is that Jan and Veronica have framed the activity differently. Veronica has framed it as "making common sense," while Jan has framed it as an "apply formal knowledge" activity. By this, she seems to mean rely on a technical vocabulary and don't worry too much about making sense of what's going on. This is illustrated in figure 14. This view is confirmed by other tutorial videos of this group, by Jan's written homework, and by six hours of interviews with Jan. Those interviews show that Jan consistently frames situations in a physics class in terms of formal knowledge that does not link to common sense intuitions. [74] They also clearly demonstrate that Jan possesses the ability to do common sense reasoning that Veronica does so effectively in Tutorial. But she doesn't feel that it is appropriate in this situation.* This dramatically illustrates that students' difficulties in doing physics the way we want them to may not be in whatever conceptual difficulties or misconceptions they possess but in inappropriate epistemological framing of a learning situation.

---

\* In the terms expressed by the authors in [74], Jan has created a "wall" between her everyday knowledge and her physics knowledge.

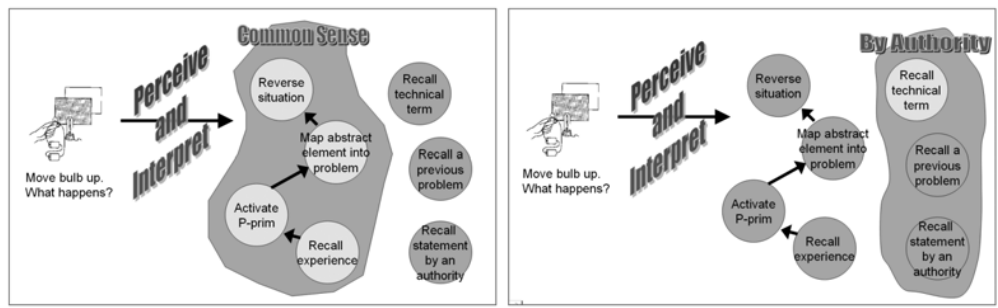

*Fig. 14: Framing of behavior models by Veronica (left) and Jan (right) in response to a tutorial activity.*

### 5.2.2    Example: E-frames can be robust and hard to shift

A second example of difficulties caused in part by being in the wrong e-frame is nicely provided by the work of Mestre and Koch as presented at this school and discussed in Mestre's paper in this volume. [61] In this work, a class of physics students was presented with the "two track" problem shown in part (a) of figure 15. Students (especially those with some physics training) tend to expect the balls to reach the end at the same time. Mestre and Koch have done a nice analysis of the resources the students appear to be using in order to come to this conclusion using p-prims and coordination classes.

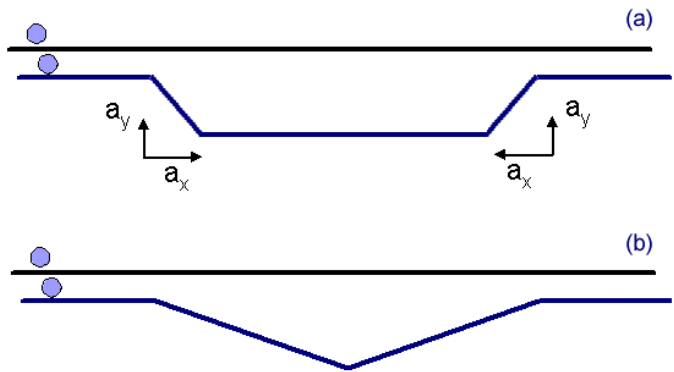

*Fig. 15: The two versions of the "two-track problem" used by Mestre. [61] Balls are launched along each pair of tracks with equal velocities and the students are asked which ball reaches the end first.*

During Mestre's presentation of this result at the summer school, I noticed an additional interesting point. Mestre presented the audience with the problem and had us discuss it in small groups and make a prediction. (I had seen the problem before. I use it regularly with my own students, so I kept silent.) One physicist in my group made the popular incorrect prediction that the balls would reach the end of the track at the same time. Mestre drew some results from the group and then explained the answer by analyzing the accelerations on the tilted parts of the lower track into components. Knowing the relation of acceleration to force, one can easily see that the pattern of accelerations leads to the lower ball always having the horizontal component of its velocity greater than the upper ball's horizontal component of velocity. It is then obvious that since the lower ball is always moving horizontally at least as fast as the upper, and sometimes faster, that it will get to the end first. My colleague smacked her forehead and said "of course!" When the second problem was presented, she solved it correctly immediately.

However, when working with a class of students, Mestre reported that presenting the analysis into components after the first problem did not help students with the second. His interviews suggest that the students continued to use the same kind of qualitative (p-prim) reasoning for the second problem that they used for the first and got the same wrong answer. Described in our new terms, my colleague framed the first problem as a simple qualitative one. Once Mestre activated a new epistemic form for her (vector component analysis), she was easily able to bring the appropriate tools to bear on the second problem. The students, for whatever reason – lack of comfort with the e-form of vector analysis, perhaps – were stuck and could not shift to a more appropriate e-frame. This suggests that there might be productive research in exploring what knowledge and associations are needed for a student to construct and activate a particular e-frame effectively.

### 5.2.3 Example: E-frames can be labile and can manage a group interaction

Considering epistemological frames can provide insight into the discourse of a group in a knowledge-building activity. A nice example is provided in the work of Tuminaro. [75] A group of students in algebra-based college physics are observed on videotape working out the solution to the problem shown in figure 16.

Three electric charges are placed as shown in the figure. Charges $Q_1$ and $Q_2$ are fixed in place. Charge $Q_3$ is free to move. As a result of the electric forces exerted on it from charges $Q_1$ and $Q_2$, charge $Q_3$ feels no net electric force. If the charge on $Q_2$ is q, what is the charge on $Q_1$?

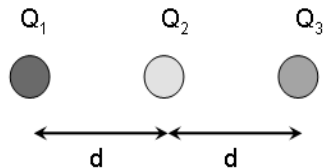

*Fig. 16: A problem in algebra-based college physics. [75]*

1.   *Joyce: Well, the distance between $Q_1$ and $Q_2$ and $Q_2$ and $Q_3$, they're the same, right?*
2.   *Karen: Yeah. I'm assuming so, since they're both d.*
3.   *Joyce: [pause] I'm thinking that the charge $Q_1$ must have is negative q.*
4.   *Karen: We thought it would be twice as much, because it can't repel $Q_2$ because they're fixed. But it's repelling in such a way that it's keeping $Q_3$ there.*
5.   *Sandra: Yeah. It has to ..*
6.   *Joyce: [overlapping] Wait…*
7.   *Karen: [continuing] Like $Q_2$ is – $Q_2$ is pushing this way, or attracting, whichever. There's a certain force between $Q_2$ that's attracting, but at the same time you have $Q_1$ repelling $Q_3$.*
8.   *Joyce: How is it repelling when it's got this charge in the middle?*
9.   *Karen: Cause it's still acting. Like if it's bigger than $Q_2$, it can still… because they're fixed. This isn't going to move to its equilibrium point. So it could be being pushed this way.*
10.  *Joyce: [quickly] Oh, I see what you're saying.*
11.  *Karen: Or pulled. You know, it could be being pulled more, but it's not moving.*
12.  *Joyce: [non-committally] Mm-hmm.*
13.  *Karen: So, we – we were thinking it was like … negative 2q or something like that.*
14.  *Sandra: Yeah. Cause it has to be like big enough to push away.*
15.  *Joyce: Push away $Q_3$.*
16.  *Sandra: Yeah. Which we – which I figured out*
17.  *Carlos: [overlapping] Yeah. I've common sense that…*
18.  *Sandra: [continuing] negative two…*
19.  *Joyce: Cause it's twice the distance away that $Q_2$ is?*
20.  *Sandra: Yeah.*
21.  *Joyce: [confidently] I agree with that.*

This conversation is interesting from a number of points of view. First, the students make a number of statements that seem quite peculiar taken at face value, but which become clear when considered in the light of the p-prim theory of reasoning resources. In line 8, Joyce remarks, "How is it (the charge on the extreme left) repelling (the charge on the right) when it's got this charge in the middle?" If Joyce knows Coulomb's law (and she does, as appears later in the exchange), why should she think that a charge in between $Q_1$ and $Q_3$ has any effect on the forces they exert on each other? It seems clear that at this point in the conversation she has not activated the epistemic game of "making meaning with equations." Rather, at this point, she and the rest of her co-workers are *making common sense* – using a combination of simple p-prims to try to generate a conclusion. In line 8, Joyce has activated *blocking* – something is in the way of an action and can prevent it. She doesn't realize that this is inappropriate since electric forces aren't blocked. Interestingly enough, in line 9, Karen does not invoke the formal rules for adding electric forces. Rather she invokes another p-prim – *overcoming*. She states that it (the charge on the extreme left) could still have an effect because it could be bigger (than the charge in the middle). Joyce's hurried response (she says line 10 at a faster than normal pace) suggests that perhaps she hasn't really understood. In lines 13 through 21, the group makes a tentative stab at an answer: since $Q_1$ is twice as far from $Q_3$ as $Q_2$ is, to balance $Q_2$, $Q_1$ must be twice as big and opposite.

Many interesting things happen as this discussion continues and as the students (with a little assist from the TA) solve the problem. This is described in table 2.

| *Description of events* | *E-Frame* |
|---|---|
| Students make some progress using p-prims, but are confused about forces and directions. | making common sense with no visual e-forms |
| The Teaching Assistant suggests they draw a diagram so they can agree on what is happening. | introduction of a new e-form |
| They now agree on which charges are exerting which forces in which directions and settle on a factor of 2. | making common sense with vector diagrams |
| Karen, recalling a previous problem tries to get them to think using the equation (Coulomb's law). She is ignored. | attempt to change the e-frame |
| Eventually, after some additional fruitless discussion, she manages to turn their attention to using the equation and the group agrees on the correct solution to the problem. | use of formal knowledge, equations |

*Table 2: Shifting e-frames in a discussion among*
*a group of students solving the problem shown in figure 16.*

Although in our theoretical framework e-framing is a cognitive process in an individual, we see here that it is useful to describe a group interaction in terms of the overlaps of the individuals' e-frames. Thus, when all the students in the group are tossing about p-prims, we can say that the group is in an e-frame in which *making common sense* is activated. If most of the students in an entire class assume that they should be following a protocol in lab and that sense-making is irrelevant to the activity at hand, we can say that the entire class is in a *from authority* e-frame (is playing the *from authority* e-game).

The three examples in this section show that paying attention to e-framing can be helpful in describing how a student is responding to a lesson, a quiz, or a group interaction. They point out that, just as knowledge structures may be robust or labile, e-framing may also be robust or labile. The cases described here are just a beginning. Classifying student e-frames in physics and determining their character for different populations is a major research effort, but the preliminary results presented here suggest that it could be a useful activity.

*5.3  Messages and Metamessages: Negotiating the E-Frame*

The e-frame a student activates can be affected by many elements in the student's environment: the layout of the chairs in the room, a technical word in the statement of a problem, etc. What the instructor says is going to happen is of course of primary importance, as is what other students say. But it is not only what is said explicitly – the *overt message* – but items in the environment and what is actually done that carry hidden implications for telling students what to do – the *covert messages*. An instructor might say in lecture "It's really important for you to learn the concepts and to make sense of the science you are learning."  That's the overt message. But if that instructor also gives homework that only requires finding the right equation, plugging in numbers, and getting an answer, he sends a covert message that contradicts his overt message. If that instructor then only tests for those limited skills, he sends a confirming covert message that says, "What really matters is knowing the equation. Making sense of it doesn't matter."

Both throughout a class and in extended interactions among students (for example, in laboratory or in group problem solving) there are continual exchanges of both overt and covert messages that sometimes result in changing which epistemic resources being applied in a particular situation. I refer to this process as *negotiating the e-frame*.

We saw an example of a group "negotiating an e-frame" in the example above in which four students were solving the problem of three charges shown in figure 16. At one point in the discussion, the students were having difficulty communicating about what forces they were talking about – on which charge, from which charge, and in which direction. The TA suggested that they draw free-body diagrams on the board – activate a previously unused epistemic form. This resulted in a shift of the group's e-frame and they made progress towards the solution. Later, Karen made a connection the other students had missed: the previous problem had similarities to the current problem and it had required the use of formal knowledge – the dependence on the force on the distance between charges from the Coulomb's law equation. At first, the other students ignored her. But a bit later, she tried again, managed to get their attention, and after some discussion (Joyce still resisting) and a prod from the TA, the group began playing the "extract knowledge from an equation" epistemic game and solved the problem.

In a classroom situation, being aware of the predominant e-frame activated by the group, the e-frames of individuals that vary from the group, and the stability or lability of those frames can help the instructor understand what is happening in her class and formulate plans for responding to it.

When we enter a new classroom situation as an instructor, we may inherit environments and constraints that send metamessages that encourage students to frame the class in a particular way.[*]  For example, when students arrives in the classroom shown on the left in

---

[*]  This is analogous to the way the set-up of a text can encourage a reader to approach the text in a particular way. Thus, a reader picking up Larry Gonick's *The Cartoon Guide to Physics* [76] is much less likely to read the text line by line, checking each result than she is do it with a *Physical Review*

figure 17, they tend to interpret the layout as a clear metamessage about what frame to activate and in which to interpret subsequent messages. Most do not expect to interact with friends or the lecturer, most expect to take notes, few expect to think about what is being said carefully and to try to understand. In the classroom shown at the right (a SCALE-UP classroom from North Carolina State [77]), even on the first day, students will be aware that this is not a traditional classroom. They may activate a group-learning epistemic resource or a new situation resource but they are unlikely to expect a lecture.

Once students have framed a situation, depending on the breadth of their experience and the consequent robustness of their framing, they may have difficulty interpreting overt messages that violate that framing. When substituting for one of my colleagues in a large lecture class, I often tell the students that I plan to have activities that will require student engagement including thinking, evaluating, and stating their views in public. Very few students take me at my word. When I call for a vote on a question, typically only about half the students respond. It is only when I call on one of the non-respondents and ask him to explain why he was unable to decide on an answer that the students begin to take me seriously.

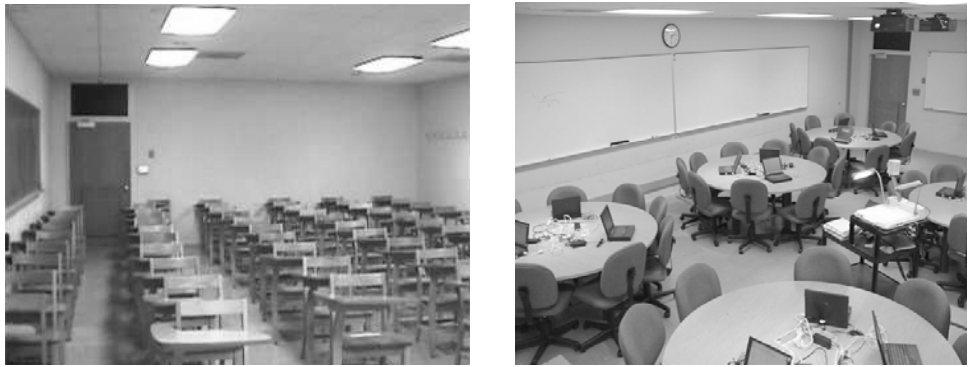

*Fig. 17: Two classrooms that send different metamessages to a student on the first day: left, a traditional lecture-style classroom, right, a SCALE-up classroom (courtesy R. Beichner).*

Reformed classes often have difficulty in "bringing the students on board." Changes often encounter massive student resistance and unhappiness. Being aware of the difference between the students' framing of our class and our own expectations can help us negotiate a frame shift with our students.

---

article. The cartoon layout sends one framing metamessage about how the text should be treated, the technical layout with abstract, references, equations, and appendices sends a different one. [69]

## 6   Implications

The theoretical frame based on neuro-, cognitive-, and social-science described in sections 2 and 3, and the resource/framing theories described in sections 4 and 5 have implications for education and educational research on a number of levels. In this section, I present a few examples of implications for instruction and for research. For some implications of this model for assessment, see Bao and Redish. [78]

### 6.1  Implications for instruction

Having a theoretical framework and some theories of how students think about physics and respond to the variety of situations they experience in a physics class can help instructors interpret what they see and help them design instructional environments more effectively. In this section, I describe some examples from my experiences working with a class of students at the University of Maryland in algebra-based (college) physics class. The class lasted for two semesters of 14 weeks each. There was some exchanging of students between my class and the class of other professors between the semesters. There were 2-3 lectures/week (with 160 students), 1 hour/week small group interaction (20 students) led by a teaching assistant, and a 2 hour/week laboratory. This class was a case-study research and development class for the NSF-supported project, "Learning to Learn Science: Physics for Biology Majors." Most of the students in the class (~85%) were biology majors and most (~80%) were college juniors and seniors (ages 20-22). Almost all (~98%) had completed calculus some years before (but had not used it much in the interval). A majority (~65%) had taken high school physics. The population was approximately evenly split by gender and most of the students were capable of mathematical reasoning but not really comfortable with it. Most expected the class to be like their other science courses – predominantly requiring memorizing many facts with protocol-guided laboratories and plug-and-chug problem solving.

As part of the NSF study, the class was extensively reformed to emphasize epistemological learning: learning how to learn science.[*] In my lectures, I adapted techniques from *Peer Instruction* [79] and *Interactive Lecture Demonstrations.* [80] Small group sections were adapted from *Tutorials in Introductory Physics.* [64] Homework problems were complex and the students were encouraged to solve them in groups in an open Course Center where TA help was available throughout the week. Laboratories emphasized measurement concepts and were not protocol based. In the lab, students worked in groups of four; they were given a brief task and had to design an experiment, carry it out, report on the results to the class, and evaluate others' lab work as well as their own. [81] I will relate some examples of how our theoretical frame has affected our work in this class in a variety of ways.

---

[*]   For descriptions and details of the various instructional methods on which these adaptations were based, see [88] and references therein.

*6.1.1    Learning to listen to students with understanding*

Over my thirty years of teaching, the way I respond to students coming to my office for help has changed dramatically. For many years, my response was reasonably consistent. The student would ask a question; I would interpret it and respond with an answer and a 5-minute mini-lecture on the topic.

An observation I made of a colleague's class when I was just beginning to get into physics education research provided a wake-up call that began to change the way I respond. I was visiting a small public university in the US when a colleague asked me to observe and comment on his lecture. I was pleased to agree. He had become a follower of physics education research and was trying to use active-engagement techniques in his large lecture class. He often called on students and tried to create some discussion (though mostly he still just lectured – but with an awareness of the student difficulties that had been described in the literature). At one point, he projected a diagram of electric field lines from three charges. The figure was quite complex with lines looping and curving everywhere. After he had finished, he called for questions. A student asked him to explain why the diagram looked like that. He treated the question as a technical one and spent a few minutes explaining why field lines never crossed. He asked if the student was satisfied and got a mumbled "Yes" in response.

I found this event quite thought provoking. Watching the student from behind and trying to read the hesitation in his voice and his body language, I felt the student was asking a much simpler question: "What's an electric field and what do those lines mean?" I was convinced that my colleague's technical answer left the student even more confused than he was at first. I was interested in the fact that I did not know which of us had interpreted the student's question correctly and I realized that often in my own response to questions I usually assumed that I understood what the student was really asking without confirming my interpretation.

Since then, I have been careful to make it my common practice to respond to a student's question with another question, one designed to clarify for me what the student is really thinking and what question the student is really asking. Often I find that a student has "technified" the question in order to make it seem he is more knowledgeable about the subject than he actually is. I often have to start any explanation I am offering at a lower level than I first expect. As I have become more experienced in questioning students, I find that questions carefully posed are often better than an explanation. If the student generates the explanation himself, he feels better about the physics and is more likely to feel that he can possibly answer some questions by himself.

As I began to learn more about the theory of student thinking, I found that student answers that had previously seemed bizarre to me (and that provided me little or no guidance for where to go next) now seemed perfectly reasonable. Using my theoretical knowledge, I am now more likely to come up with an appropriate response. Let me give you an example.

About the middle of the first semester, soon after our study of energy, one of my students came to my office hours with a complaint. "I understand energy conservation, but I get all the problems wrong!" In order to try to get some idea of what was wrong, I responded, "OK. Let's consider the following problem. I'm holding this eraser. If I drop it, is its energy conserved?" She responded with a confident and dramatic, "No!" Taken somewhat aback by her confidence and apparent lack of confusion in delivering the wrong answer, I responded in a way my students had become accustomed to hearing, "OK. Tell me why you say that." She responded, "Well, it starts out with a large potential energy and as it falls, that gets less!" This removed my concern that she didn't understand potential energy and was only thinking about kinetic. But what was she doing? I probed further. "And what about the kinetic energy?" She responded, again confidently and vigorously, "That changes too!"

Now that was a really bizarre response from a physics point of view. She clearly knew what both potential and kinetic energies were and how they both changed. Why didn't she see the conservation? My understanding of the resources we had seen our students use in other physics problems came to my aid. Jonathan Tuminaro identifies a resource (primitive) he calls *feature analysis*. This is a resource used in everyday thinking to compare objects and decide whether they are equivalent. An example is shown in figure 18. Two cartoon faces are drawn. Are they the same? Checking the eyes shows they are a bit different. Is this enough? A look at the nose shows they are different too. Putting the two results together is more convincing: they <u>are</u> different. I refer to this resource as *different plus different is more different*. This is the resource my student was activating rather than the *compensation* resource I had expected her to use.

But *compensation* is clearly a resource that I can expect every college student to have for at least some contexts. If I could find a bridging analogy [82] that would activate the appropriate resource it might help her. I suggested she consider that she was holding a stack of 20 one-dollar bills and that she start handing them to me. As she began to go through the motions I asked her, "Is the number of bills you have conserved?" She responded without hesitation. "No." Is the number of bills I have conserved?" I continued. She again confidently responded, "No." I began, "Is the total number of bills in the room…" and she gasped, looked up, put her hand to her mouth and said, "Ooooh!" Once she had activated compensation instead of the competing feature analysis she was able to make sense of the energy conservation situation and answer three or four more similar problems without difficulty.

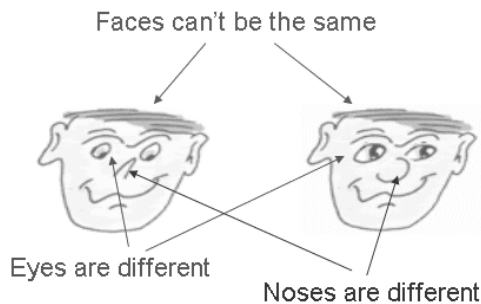

Fig. 18: *Feature analysis. Are the two faces the same? A comparison of features gives the answer. The more features that differ, the clearer it is that the two are different.* [75]

I am not suggesting that I would not have been able to come up with an appropriate bridging analysis without the resource theory. I hope I would have. But I am quite certain that had I not thought about and seen students apply many different everyday resources to physics problems – both correctly and incorrectly – that I would have been much more surprised at her "That changes too!" and been very confused as to what her problem was. The theory helped me make sense of what I was seeing and changed my first judgment of her response from stupid/bizarre to "perfectly natural."

### 6.1.2    Situating cognition: Coherence

The previous example shows how just being aware of the kinds of resources students have and that they might or might not activate them appropriately can help you understand some of what your students say. Associational patterns are also of great importance.

A cognitive instructional model that pays a lot of attention to what we would call associational patterns in our theoretical framework is the theory of *situated cognition*. [83] This model relies on the idea that in authentic activities in our daily lives we build up substantial associational patterns of resources that allow us to solve many complex problems – as long as they appear in the appropriate context.

An example from cognitive research is the problem given by Wason and adapted here in figure 19. [84]

A label machine cuts labels and prints a letter on one side (either an A or a B) and a number of the other (either a 2 or a 3). It has no other characters and never prints anything but these and on the proper side. However, on every label that has an "A" it is supposed to print a "2" on the other side. Sometimes it slips, however, and makes a mistake on this.

If you are a checker checking labels as they come past you on the assembly line out of the machine, which of the following labels flowing past you would you turn over to check that they are done correctly?

| A | 3 | B | 2 |

*Fig. 19: A difficult problem. [84]*

This problem is quite difficult, and most people miss it. However, if the problem is situated in a context that activates appropriate resources, people can solve it much more easily. A more appropriately situated version is shown in figure 20. Using their social and cultural knowledge, most adults can solve this problem with no difficulty.

Both problems are equivalent and have the structure *p →q is equivalent to ~q →~p*. However, the ability to solve the situated version of the Wason problem does not help most people solve the more abstract form. The use of compiled, easily activated context knowledge to solve situated cognition problems is all very well, but it is not our goal. We want to establish more general solution patterns that work in a wide variety of contexts. How can we make use of compiled situated context knowledge but facilitate far transfer, making it useful in a wider variety of contexts?

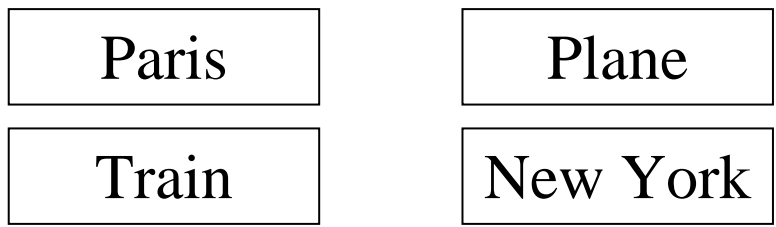

You are managing a group of students who have planned summer research travel from your university in Milan. If their assignment is in Europe, they are supposed to go by train to save money. If it is in the USA, they go by plane.

Your secretary gives you a set of cards, one for each student. On one side is indicated the student's destination, on the other whether the student is scheduled to go by train or plane. Which cards do you need to turn over to make sure no students with European destinations have scheduled to fly?

| Paris | Plane |
| Train | New York |

*Fig. 20: A problem one easily solves using situated knowledge.*

One solution that naturally occurs to most physicists is to go abstract. Anyone who knows Boolean algebra should realize $p \rightarrow q \leftrightarrow \sim q \rightarrow \sim p$ and be able to easily solve both problems. However, as we have seen, knowing something and activating it in the proper context is not as easy as it sounds.

A nice example of this was given at this school by Mestre in his talk about problem posing. [61] In an interview, he gave a good student in calculus-based physics the task of creating "end-of-chapter textbook problems" for particular physical situations. In one case, the situation was the "half-Atwood's machine" as shown in figure 21. One of the problems posed by the student for this situation was to find the acceleration of the system if both masses were equal. The student sketched out a solution but incorrectly identified the tension as being $m_1g$. This led to the two masses having different accelerations, a situation that troubled the student. His intuition was that if the masses were equal, the acceleration should be ½g. (This is good p-prim reasoning!) But in the end, he rejected his intuition and said, "Math doesn't lie!"

I have seen a similar situation with many of my students. They do not seem aware that the *mapping* of the situation into math could be wrong. Once they have translated the problem into an abstract equation, they assume it has to be right. This undermines our attempt to use situated cognition as a basis for establishing correct reasoning.

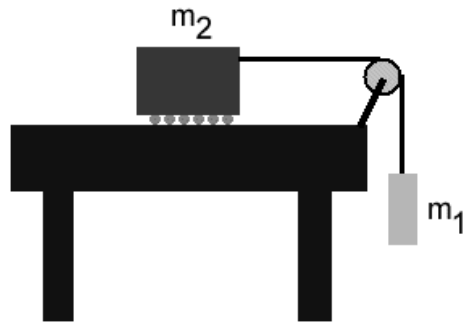

*Fig. 21: The half-Atwood machine used by Mestre in problem posing. [61]*

If we want to make use of the good intuitions our students have – and help them decide when their intuitions are good and when they're not – we have to help them to explicitly make connections between the abstract and formal reasoning they are learning and their everyday experience. I refer to this process as *situating cognition*. One way I do this is with homework problems that explicitly ask students to make the connection to their everyday experience. An example is shown in figure 22.

One day I was coming home late from work and stopped to pick up a pizza for dinner. I put the pizza box on the dashboard of my car and pushed it forward against the windshield and left against the steering wheel to prevent it from falling. (See picture.) Before I started driving, I realized that it could still slide to the right or back towards the seat. When driving, do I have to worry about it sliding more when I turn left or when I turn right? Do I have to worry more when I speed up or when I slow down? Explain your answer in terms of the physics you have learned.

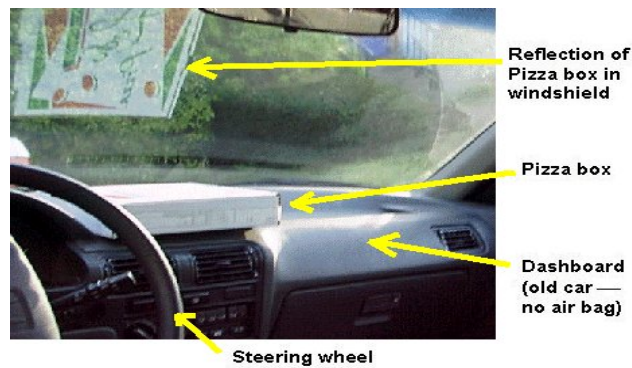

*Fig. 22: A problem to help students situate Newton's laws in their everyday experience. [85]*

### 6.1.3    *To Conflict or Not to Conflict: Reconciling*

Traditional reform methods have relied heavily on the cognitive conflict technique when students map their everyday experience into a physics class incorrectly. [86] [87] [88] Students are asked to make a prediction as to what will happen in a particular situation, chosen because it is known that a majority of students will make an incorrect prediction. They are then asked to observe an experiment that contradicts their prediction and are helped, usually, to understand why it doesn't work the way they thought it did, but not to resolve the conflict with their original idea.

From our theoretical framework, we see two dangers to this approach.

1.  The students may create an independent knowledge element without modifying their original knowledge element.
2.  The students may be led to cut their experience and physical intuition out of the frame they activate for doing physics.

The first seems to be what Jan did in the example discussed in section 5.2.1. The second seems to be what Mestre's student did in the example discussed in section 6.1.2. Hammer cites the case of one of his students who, upon receiving a no-grade pre-test as the first step of one of the class's cognitive conflict lessons, sighed and said, "Here's another quiz to show me how stupid I am about physics."

Our theoretical structure's emphasis on associations and epistemology highlights the dangers with an approach that, on the surface, produces impressive gains on limited measures (conceptual gains on standardized tests).[*] We are led to ask: Can we retain the conceptual gains produced by the best cognitive conflict methods without doing damage to the student's epistemological framing?

One method, developed by Andrew Elby, works to shift the method from a conflict between the student and authority-taught physics to a conflict within the student himself. Elby relies heavily on his knowledge of physics resources to predict how students will respond to questions formulated in particular ways. Since resource application is highly context dependent, Elby is able to create pairs of questions (*Elby pairs*) that ask the same question in two different ways. In one formulation, the student is likely to answer the question with a common misconception.[†] Students usually get this wrong. In Elby's second formulation, the question is formulated so as to match the student's intuition. Since students have navigated the physical world successfully for more than a decade, their intuitions usually have a thread of truth. (For example, students who believe that heavier objects fall faster than light objects are inappropriately generalizing their correct intuitions about falling rocks and feathers.) Students usually get this second question right. By then showing students that their two intuitions are in conflict, he helps them to reconcile their own internal knowledge with the physics they are learning. As a result, the students not only learn the physics, but how to relate it to their intuitions. They learn to check, evolve, and rely on their intuitive physics.

---

[*]   See for example the discussion of various instructional methods in [88].

[†]   There are large collections of such questions in many areas of physics developed by researchers and by curriculum developers who use cognitive conflict. See, for example, [79] and the Action Research Kit on the CD associated with [88].

A classic example of this technique is given in Elby's paper introducing the Elby pair. [89] Students are well known to have difficulty believing in Newton's third law. Although they can often state the law (especially the "action-reaction" form), they often either don't know what the words mean or don't believe that the law applies widely. Elby poses the pair of questions shown in figure 23.

A heavy truck rams into a small parked car.
    1. Intuitively, which is larger during the collision:
        the force exerted by the truck on the car,
        or the force exerted by the car on the truck?
    2. Suppose the truck has mass 1000 kg and the car has mass 500 kg.
        During the collision, suppose the truck loses 5 m/s of speed.
        Keeping in mind that the car is half as heavy as the truck,
        how much speed does the car gain during the collision?
        Visualize the situation, and trust your instincts.

*Fig. 23: An "Elby pair" of questions on Newton's third law.*

Approximately 75-90% of my students will miss the first question, even after instruction. But question 2 is correctly answered by 90% of my students. Elby has phrased the question so as to activate a proportional scaling p-prim. We then analyze the implication of the two answers through a reconciliation diagram such as the one shown in figure 24. The first step is to identify the core or "raw" intuition that leads to both results. In the case of the third law problems, many students will respond that "the car reacts much more than the truck does." When the students figure out that the Newton's second law allows the car to accelerate twice as much as the truck, even when the forces are the same, they are much more willing to accept the idea of Newton's third law. When their refined intuition is confirmed by an experiment, Newton's third law doesn't just become something they have to accept, it's something that makes sense. In my classes, using this technique led to fractional gains on the Newton's third law cluster of question on the Force Concept Inventory comparable to those obtained with cognitive conflict Tutorials (~75-85%) but without the loss of epistemological sophistication typically found in introductory physics.[*] Creating Elby pairs is greatly facilitated by the resource theory based on p-prims and modular responses.

---

[*] For gains on the FCI Newton 3 cluster arising from cognitive conflict Tutorials, see [90]. Epistemological sophistication is measured by the Maryland Physics Expectations Survey (MPEX). [91] This is a broad survey, so it is not responding solely to this lesson, but to the whole package of epistemological modifications. Typical classes show pre-to-post losses on the MPEX, even reformed classes that produce good conceptual gains. Our modified classes show a strong gain on this measure.

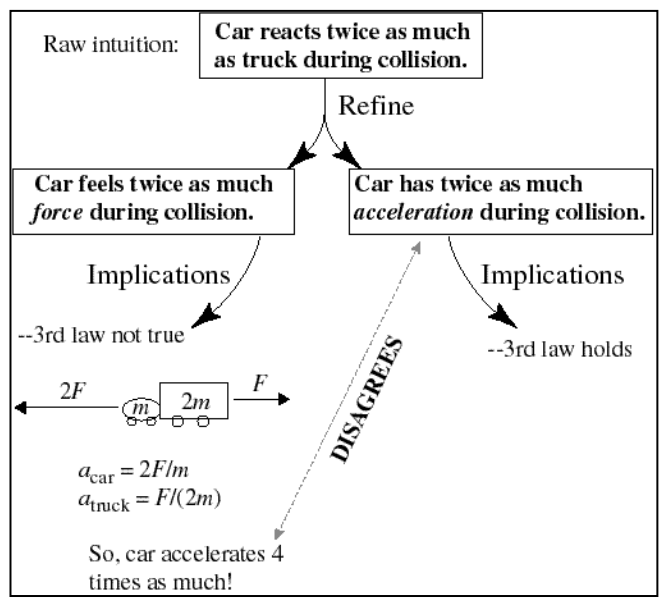

*Fig. 24: An Elby reconciliation diagram resulting from the discussion with students of their responses to an Elby pair of questions. [89]*

## 6.2  Implications for research

Besides having implications for instruction, working with a theoretical framework such as the one described in this paper can have significant impact on our research as well. It provides us with an understanding of the range of possible interpretations for our observations of student's behavior. To illustrate this point, I describe two examples from the research of the group at the University of Washington and one from the research of the group at the University of Maryland.

The group at the University of Washington has a long history of studying student difficulties with learning physics at the college level. They have studied topics ranging from the concept of density and mass to relativity and wave functions in quantum mechanics. They have documented many specific student difficulties, some of which are interesting to analyze from a theoretical point of view.

One such case that was discussed by Heron at her talks in this school occurs in Ortiz's study of student difficulties with torque and the statics of extended objects. [92] After instruction, students were shown a picture of an irregular wooden object (a baseball bat) on which the center of mass was marked and a finger was shown balancing the bat above that point as shown in figure 25. Most students stated without hesitation that the masses on both sides were equal. When asked, they made statements about balancing, some mentioned torque, and one student said, "That's what center of mass means."

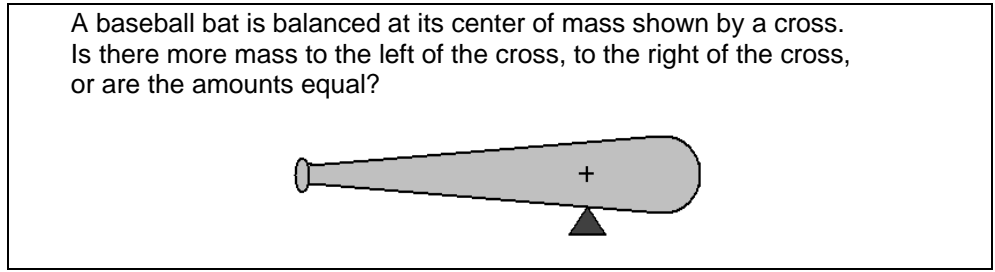

A baseball bat is balanced at its center of mass shown by a cross. Is there more mass to the left of the cross, to the right of the cross, or are the amounts equal?

*Fig. 25: A problem posed by Ortiz and*
*the Physics Education Group at the University of Washington. [92].*

This raises some interesting questions. How should we interpret this result? Do we believe that the students do not understand torque and the balance condition? We might then try to remedy the difficulty by specific remedial instruction on torque. But what if the issue is a framing one? If the students were not yet fully comfortable (had not yet compiled) the ideas and procedures for using torque, they may have framed this problem as an easy and "obvious" one – a problem where they didn't need to activate the "hard" physics principles but where a p-prim would do just fine. It seems very likely that students would activate *balancing* in this context and, cued by the phrase "center of mass," map "mass" onto what is balanced. How would we know?

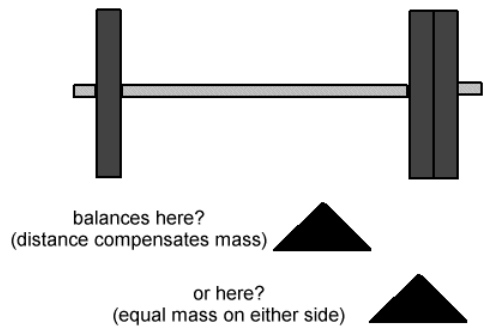

balances here?
(distance compensates mass)

or here?
(equal mass on either side)

*Fig. 26: Extension of the baseball bat problem to test*
*the organization of student knowledge structures.*

If the students really thought that the balance point had equal masses on both sides, we could test it by posing them a barbell problem – two heavy disks attached through their center by a long, light (and strong) rod – in which one disk is twice as thick and heavy as the other. If the students believed the balance point principle, we would expect them to say that to balance the barbells, you would have to put your fulcrum under the wider disk. (This is illustrated in figure 26.) This seems to me unlikely. I would expect that they would activate *compensation* and say, "If the right disk is twice as heavy, you have to balance at a point twice as close to the heavy disk." Such responses in a significant fraction of the target population would provide strong support for students having a modular knowledge structure at this stage of their learning this topic.

A second nice example mentioned by Heron comes from the research of Loverude. [93] [94] Students were asked to consider two identical graduated cylinders. Two heavy balls of the same size but different masses were put in the cylinders and the cylinders filled with equal amounts of water. The students were asked in which cylinder the water would rise higher. Many students responded that it would rise higher in the cylinder with the heavier ball. Why did they do that? The cylinder problem seems trivial on the surface. Failure here might signify a serious problem with the concept of volume, mass, or displacement. This might be true for some students. But we might consider some other options suggested by our theory. Depending on exactly how the question is presented, there could be a number of cues that could lead to students framing this as a "trivial" question – one that can be answered with exam-solving skills rather than with real-world thinking. By the time they reach college (especially in the US where high school students take frequent multiple choice tests under time pressure), students have had considerable experience performing "triage" on a test – finding obvious or easy questions and quickly getting them out of the way. If the students had framed this question in that way, the easiest way to solve it is to activate a *more is more* p-prim and map it as "more mass" means "more displacement." As in the first example, having an explicit theoretical framework makes it less likely that we are implicitly using a theory as a "hidden assumption" without considering a wider range of possibilities.

Both of these examples show that when we are doing research on student difficulties, having a theory of the way students think can help formulate competing hypotheses about what the students are doing when they give wrong answers – hypotheses that we might otherwise miss. Having a variety of possibilities helps formulate questions that can further guide the research, help us understand how to improve our models of student thinking, and help us understand how to create effective lessons.

A third example comes from the research of Wittmann while he was at the University of Maryland. As part of a research project to study students' concept of electric current, he interviewed a physics major in her junior year, code-named Sarah. [95] In the interview, Wittmann was attempting to explore Sarah's conceptual knowledge. In a subsequent analysis of this interview, Wittmann and Scherr noted that the interviewer's framing of the interview as a probe of conceptual knowledge resulted in his ignoring some interesting information about Sarah's chosen sources of knowledge. [96] From the point of view of obtaining information about Sarah's conceptual knowledge, the interview begins slowly. Sarah seems to have initially framed the interview as a test; she is quite concerned about the correctness of the answers she is giving and frequently cites authorities (other classes, memorized information, texts). As the interview progresses, the interviewer provides some suggestions, and, at a critical point, she seems to reframe the interview, beginning to move into an e-frame that activates *knowledge as fabricated stuff*. She uses other ideas and intuitions in new ways and combinations ("That's stuff I've never thought about before, and I'm just making up as we go along!") to answer the questions in a way that makes sense to her. Had we better understood the nature of epistemological framing eight years ago, we could have substantially improved Wittmann's understanding of what was happening in his interview at it occurred.

My examples, the office-hours interaction, the development of effective tutorial lessons, and understanding the subtleties of what is happening in a research interview, do not present results that are new or that have not been achieved by others. The best teachers, curriculum designers, and interviewers often have highly developed skills that allow them to effectively accomplish their goals. But learning those skills is often extremely difficult and can take years as an apprentice working with a master teacher or researcher. Our goal of "trying to study education scientifically" is to find ways to create many more master teachers, curriculum developers, and researchers more quickly. Having a way to describe and explain what is going on in these interactions has proven to be an excellent way of doing this in a variety of fields from art to engineering.

## 7  Conclusions

In this paper, I have begun to outline a theory for studying the teaching and learning of physics and have described some models that fit within it. The framework is at present focused on the individual, how the individual thinks and how the individual's thinking interacts with her environment. As the models that fit in this theory become better established, they must be interfaced and meshed with theories of social and cultural interaction.

### 7.1  Summary

The theoretical framework discussed here is based on structures uncovered in neuroscience and cognitive science and the interaction between them. It consists of intermediate-level cognitive structures (resources) that are mesoscopic (complex when viewed from the neural level but simple when viewed from the level of individual thinking and behaving). Broadly, I focus on two sets of structures: associations and controls. Models within this framework identify resources, their associational patterns and their context dependence. Associational patterns identify which resources go together and controls identify the environments and cues that activate particular associational patterns.

Some of the axes along which models have to be defined include: the robustness of the associations (probability of activation in appropriate circumstances), their degree of compilation (extent to which they can serve as a single unit in working memory), and their degree of integration with other related elements. Within the two structures of association and control, I identify some models appropriate for analyzing student thinking and learning about physics.

In the category of association, I identify phenomenological primitives as fundamental resources and refine the idea slightly to separate them into abstract reasoning primitives (of which there are few) that can be mapped into facets describing specific phenomena (of which there are many). [48] This splitting offers some advantage in developing instructional environments to help students transform "misconceptions" that are basically inappropriate mappings. One associational structure of resources is discussed here, the coordination class. [49] [59] Many more need to be developed.

In the category of control, I adapt the idea of frames and framing from the behavioral science literature. I pay particular attention to learning – the construction of new knowledge and the solution of problems. This requires a variety of epistemic resources that are organized into associational patterns called games and forms. [65] The control structures that activate particular patterns of games and forms are called epistemological frames.

We then offer a number of examples showing how even at this crude level of development, thinking in terms of these theoretical structures can help diagnose student difficulties, create new lessons and instructional environments, and clarify issues in research at a number of grain sizes ranging from guiding an interview to comparing different theories of learning. Although there is much to be done, having a theoretical framework helps on a variety of levels.

### 7.2  Speculations

I conclude this paper with two speculations.  First, recall that in section 4.2, I compared two common models of associational patterns in student resources in thinking about physics that I referred to as the modular and model models. In some ways student knowledge systems are disconnected, labile, and chaotic. In other ways it ways, they are rigid and robust. I quoted an example from the work of Scherr et al. in which students showed characteristics of both models. In the study of complex systems (see for example [97] and references therein) a number of examples have been identified where a system can "ride the boundary between chaos and stability." The phenomenon is called *self-organizing criticality*. There is a certain plausibility to the idea that evolution could drive the development of a cognitive system that has a certain degree of stability in order to be able to create predictions and a certain degree of flexibility in order to be able to respond to novel situations. Thus, it might be appropriate for us to seek theories of resource association that are first and foremost, neither model-based nor modular but something that balances between them. What that theory would look like, I don't know.

A second point that is potentially of great importance is *affect*. By this I mean a combination of emotional responses, motivations, and self image. Every teacher knows these issues are of great importance in a student's learning or not learning. The cognitive-neuroscience interface seems to indicate that there is a significant interplay between emotion and rational reasoning. [9] But if anything, emotion is a even a more complex phenomena to study than cognition since it involves not only local neural processes but more systemic chemical and glandular phenomena as well. At this stage of research the system appears to me to be not as well understood as the more direct cognitive processes. Nonetheless, it is clear that affect phenomena are of great importance and eventually should be a third branch of our theoretical frame, strongly interwoven with association and control.

# 8   Glossary

One of the difficulties with exploring the education and cognitive literatures is an inconsistent use of terms. Commonly used terms such as "schemas" or "mental models" are used with different meanings by different authors leading to considerable confusion. Often, terms are not defined at all, even by citations to the literature. To try to avoid this pitfall, and because this paper introduces many terms that may be unfamiliar to my readers, I include a glossary.

*Activation* – the level of activity of a neuron or set of neurons. When a neuron is activated it sends a chain of electrical pulses (action potentials) down its axon. A neuron can be in a variety of activation levels.

*Association* – two resources (or neurons) are said to be associated when the activation of one leads to the activation of the other in some context or set of contexts.

*Axon* – a long thin protuberance extending out from a neuron (*q.v.*). The axon carries electrical pulses. The frequency of these pulses carry information.

*Belief* – an articulated statement about something or about a state of affairs that affects a variety of behaviors relating to that subject.

*Causal net* – an ordered associational pattern of resources.

*Chunking* – the tight association of multiple resources or knowledge elements so that they can be used as a unit in working memory. (cf. compilation)

*Cognitive conflict* – an instructional method in which a student's belief or misconception is first activated (often by calling for a prediction) and then challenged by evidence (often experimental).

*Compilation* – the process by which a group of related knowledge elements become tightly associated (perhaps through familiarity and practice) so that they activate reliably and easily together and can be used as a unit in working memory. (cf. chunking)

*Constructivism* – the belief, common among educational researchers today, that new knowledge must be constructed out of existing knowledge, by establishment of new associations, transformation, and processing.

*Context* – in common speech, the environment in which a task is presented; in our theory, the complete activation state of the neurons in a brain when a task is presented to it. To understand context, one must decide what parts of those activations are relevant to the task at hand and how those activations influence control elements.

*Control* – the process by which particular resources are selected for activation instead of other resources that might be relevant. Can involve both activation of some resources and inhibition of others.

*Coordination class* – an associational pattern of resources involving a readout strategy (*q.v.*) and a causal net (*q.v.*).

*Covert message* – a message that is unintentionally or tacitly sent by a speaker or environment, especially one that leads a listener to frame a situation in a particular way. (*cf.* overt message)

*Elby pair* – a pair of questions that ask the same physics question in two different ways. In one way, the context of the question cues a common student misconception with a high probability. In the second way, a different context cues a correct response. The pair of questions create a "teachable moment" to help the students reconcile the two responses and refine their intuitions to be more in line with accepted scientific descriptions.

*Epistemic form* – the cognitive tools to manipulate and interpret an external structure or representation useful for building knowledge or solving problems. Often (by metonymy), the external structure itself.

*Epistemic game* – a coherent activity that uses particular kinds of knowledge and the processes associated with that knowledge to create knowledge or solve a problem.

*Epistemic resource* – a schema (*q.v.*) used for constructing knowledge. For examples, see epistemic game and epistemic form.

*Epistemological frame* – the set of epistemic resources the individual assumes is appropriate to carry out the task at hand.

*Facet* – a specific statement describing the functioning of a particular phenomenon or system.

*Framing* – the process of selecting a subset of information in the environment as relevant and the associated activation of appropriate resources to deal with it.

*Fine-grained constructivism* – an approach to learning that in a given learning situation asks, if new knowledge must be built from old (cf. constructivism), what particular knowledge is used and how is it transformed?

*Foothold idea* – an assumption that is maintained as a working hypothesis.

*Lesion studies* – the study of how an individual's cognitive processes change as a result of particular limited brain damage (usually obtained from stroke or injury). Gives information on the extent to which particular functions can be localized in the brain.

*Links* – associations (q.v.) between knowledge elements or resources.

*Long-term memory* – the part of memory in which knowledge and information are stored in a stable fashion. Long-term memory has immense capacity but it may take substantial effort to create an item in long-term memory and it may take substantial effort or time to activate it.

*Mapping* – the association of the indeterminate elements in a schema or reasoning primitive with specific elements of the world.

*Mental model* – an association pattern of cognitive elements that fit together to represent something. Typically one uses a model to reason with or calculate from by mentally manipulating the parts of the model in order to solve some problem.

*Metamessage* – a message sent by a speaker or element in the environment that leads a listener to frame subsequent messages in a particular way.

*Misconception* – a knowledge structure that is activated in a wide variety of contexts, is stable and resistant to change, and that disagrees with accepted scientific knowledge.

*Model* – This term is used in a number of senses. In the physics community, it refers to a set of coherent assumptions that are intended to explain and help understand a fairly limited class of phenomena. It may be held only by a few (or by a single) researchers. Such structures are often called theories in education. See also mental model.

*Model model* – the model of student thinking (about something – such as a topic in physics) that posits that student thinking consists of well integrated coherent and consistent schemas that are used in a wide variety of contexts.

*Modular model* – the model of student thinking (about something – such as a topic in physics) that posits that student thinking largely consists of weakly associated schemas that are highly context dependent and poorly integrated.

*Neuron* – specialized cells in animals for the collection and processing of sensory data, for thinking, and for initiating and controlling muscular activity.

*Ontology* – the description of a system in terms of the kinds of objects relevant for its description and their characteristics.

*Overt message* – what is said explicitly, in particular, a message that is intended to help the listener frame the conversation. (*cf*. covert message)

*Pattern of association* – the spontaneous activation of different resources or knowledge elements in conjunction with each other in a particular context.

*Phenomenological primitive* – a cognitive resource corresponding to a basic statement about the functioning of the physical world that a user considers obvious and irreducible.

*P-prim* – a phenomenological primitive.

*Priming* – the partial or low-level activation of a set of resources by a particular input. Once resources are primed they are typically easier and quicker to access than if they are not primed.

*Readout strategy* – a resource for the interpretation of sensory input in a coordination class (*q.v.*).

*Reasoning primitive* – an abstract reasoning principle such as "more cause leads to more effect." When mapped (*q.v.*) into a physical system it becomes a facet (*q.v.*).

*Reductionism* – description of a system in terms of the behavior of constituent parts and their interactions.

*Resource* – a compiled knowledge element, typically one that appears irreducible to the user. Since different individuals have compiled their knowledge in different ways, different levels of structure may be used as resources by different individuals.

*Schema* – a bounded, distinct, unitary cognitive representation or pattern of association that is not too large to hold in working memory.

*Spreading activation* – the sequential activation of a series of resources.

*Theoretical framework* –The education community's term for a community's shared and shared assumptions. Typically includes both ontological elements (basic components and their interactions) and epistemological elements (how one decides one knows something). Usually a t.f. is meant to cover a wide range of phenomena. Specific theories (*q.v.*) or models are required in order to generate predictions and explanations in specific cases. Equivalent to what is called a "theory" in physics.

*Theory* –In the physics community, this refers to a broad general structure of coherent assumptions widely believed by a scientific community. In the education community, it refers to a set of coherent assumptions that are intended to explain and help understand a fairly limited class of phenomena. It may be held only by a few (or by a single) researchers. The education community's use of the term is approximately equivalent to the physics community's use of the word model (*q.v.*).

*Working memory* – the part of memory in which knowledge and information are mixed in active thought. Working memory has a limited capacity and items in working memory fade and deactivate if not restored by a process such as rehearsal (conscious repetition) or chained activation from other elements.

*Zero-friction experiment* – an experiment that is carried out to probe a fundamental mechanism by suppressing or controlling elements that are typically present in everyday examples of the phenomenon being studied.

## 9   Acknowledgements

This work is supported in part by NSF grants REC-0087519 and PHY-0244781. I gratefully acknowledge many stimulating discussions with my colleagues at the Varenna summer school, especially Andrea diSessa, Noah Finkelstein, Jose Mestre, and Valerie Otero. This work has been greatly affected by many valuable interactions with my colleagues at Maryland, particularly Andy Elby, David Hammer, Rebecca Lippmann, Laura Lising, Rachel

Scherr, and Jonathan Tuminaro. Finally, I gratefully acknowledge the many conversations on this subject that I have had with my wife, Janice (Ginny) Redish. Her insights into language and human behavior have been invaluable.